\documentclass[10pt,twocolumn]{article}

\usepackage[utf8]{inputenc}
\usepackage[T1]{fontenc}
\usepackage{times}
\usepackage{geometry}
\usepackage{authblk}
\usepackage{abstract}
\usepackage{titlesec}
\usepackage{setspace}
\usepackage[hidelinks]{hyperref}
\usepackage{graphicx}
\usepackage{amssymb}
\usepackage{tabularx}
\usepackage{array}
\usepackage{comment}
\usepackage{amsmath}
\usepackage{comment}
\usepackage{moreverb}
\usepackage{graphicx}
\usepackage{stfloats}
\usepackage{graphicx}
\usepackage{subcaption}

\usepackage{enumitem}
\usepackage{lipsum} 
\usepackage{longtable}
\usepackage{booktabs}
\usepackage{amsmath}

\usepackage{orcidlink}

\usepackage[
    backend=biber,
    style=numeric-comp,
    sorting=none,
    natbib=true
]{biblatex}

\titleformat{\section}
  {\normalfont\bfseries\uppercase}
  {\thesection.}{0.5em}{}

\titleformat{\subsection}
  {\normalfont\bfseries}
  {\thesubsection}{0.5em}{}

\title{\textbf{NeuralFlowNet: Towards Data-Free Physics-Informed Neural Network Solutions of Navier--Stokes Equations Across Low and High Reynolds Numbers}}

\author{
Jayanga T. Samarasinghe\textsuperscript{1,2,a}\,\raisebox{-0.2ex}{\scalebox{1.5}{\orcidlink{0000-0001-8491-0092}}},
Luis De la Fuente\textsuperscript{2,b}\,\raisebox{-0.2ex}{\scalebox{1.5}{\orcidlink{0000-0001-6979-0547}}},
Laura V. Alvarez\textsuperscript{1,2,3,c}\,\raisebox{-0.2ex}{\scalebox{1.5}{\orcidlink{0000-0002-5047-5384}}}
}

\date{} 

\begin{document}

\twocolumn[
\vspace{-10pt}
\maketitle

\vspace{-25pt}
\textbf{AFFILIATIONS}\\
\textsuperscript{1}Environmental Science and Engineering Program, University of Texas at El Paso, El Paso, Texas\\
\textsuperscript{2}Department of Earth, Environmental and Resource Sciences, The University of Texas at El Paso, El Paso, Texas\\
\textsuperscript{3}Institute for Applied AI Innovation, The University of Texas at El Paso, El Paso, Texas\\\vspace{6pt}

\textsuperscript{a}jthambanged@miners.utep.edu, corresponding author\\
\textsuperscript{b}ladelafuenteco@utep.edu, corresponding author\\
\textsuperscript{c}alvarez@utep.edu, corresponding author\\

\begin{onecolabstract}
Physics-informed neural networks (PINNs) have emerged as a compelling pathway toward trustworthy artificial-intelligence-based computational fluid dynamics (CFD) by embedding the governing equations directly into the learning process. This is important because many existing AI flow models require large simulation or experimental datasets and often remain problem-specific, limiting their generalization and physical reliability. Data-free PINNs frameworks offer an alternative by learning flow solutions from the Navier--Stokes equations and prescribed boundary conditions, potentially reducing dependence on expensive CFD datasets while retaining physical consistency. However, traditional PINNs frameworks have failed to solve high Reynolds numbers, which has limited the application of those AI-tools in CFD. In this work, we present \textit{NeuralFlowNet}, a data-free physics-informed proof-of-concept framework designed to solve steady Navier--Stokes problems across low- to high-Reynolds-number conditions in steady-state problems. The objective is to evaluate whether  \textit{NeuralFlowNet} can provide physically meaningful and reliable flow predictions without training on external flow-field data, while also identifying the methodological requirements and limitations of extending such models toward more complex regimes. We first describe the proposed methodological framework and then demonstrate its applicability using a sequence of benchmark problems with increasing physical and geometric complexity. The results demonstrate that \textit{NeuralFlowNet} can accurately recover steady flow fields across a broad Reynolds-number range, including cases with strong pressure gradients, while maintaining good agreement with the reference numerical solutions. These findings establish \textit{NeuralFlowNet} as a reliable framework for future testing unsteady and more complex simulations, which could set the basis for creating trustworthy and efficient AI solvers for fluid dynamics simulation with high Reynolds numbers.

\end{onecolabstract}
\vspace{0.5cm}
]

\section{Introduction}

Over the past few decades, computational fluid dynamics (CFD) has become the primary numerical framework for approximating solutions to the Navier–Stokes equations (NS) across diverse flow regimes. Methods such as finite volume, finite element, spectral, and mesh-free approaches have enabled major advances in fluid-flow simulation \parencite{brooks_streamline_1982,karniadakis_spectralhp_2013,katz_meshless_2009,kramer_lattice_2020}. These methods have been applied successfully to laminar, transitional, and turbulent flows \parencite{hafeez_review_2023}. However, several challenges still limit the practical use of conventional CFD method in real problems. For instance, in complex industrial and environmental problems, mesh generation is often time-consuming and highly problem-dependent \parencite{chawner_geometry_2016,ito_challenges_2013,inthavong_computational_2021, zhang_prototype_2010}. It also requires significant practitioner expertise and often relies on expert judgment \parencite{chawner_geometry_2016, park_unstructured_2016}. Moreover, inverse problems, where unknown boundary conditions, source terms, or physical parameters must be inferred from available observations, are even more challenging \parencite{beck_inverse_1985,das_simulated_2012,wang_recent_2024}. Such problems often require repeated forward simulations and significant computational cost. Modern CFD solvers, such as what are implemented in OpenFOAM, rely on large and specialized software infrastructures, including mesh generation, discretization schemes, solver settings, boundary-condition implementation, parallel execution, and postprocessing workflows \parencite{jasak_openfoam_2007}. Although powerful, these tools often require significant user expertise, careful case-specific configuration, and substantial computational resources. As a result, they can be difficult to maintain, modify, and extend. These challenges have motivated increasing interest in alternative computational frameworks that retain physical fidelity while offering new ways to represent and solve fluid-flow problems.

The emergence of machine learning (ML) has influenced CFD in multiple ways. Existing efforts include forward modeling, data-driven surrogate modeling, physics-driven surrogate modeling, and ML-assisted numerical solution strategies \cite{wang_recent_2024}. In addition, ML has been applied to inverse design \parencite{lu2021physics,wu2022learning,yu2022gradient}, flow control, and a broad range of fluid applications across different disciplines. Together, these developments highlight the growing role of ML in the modeling, analysis, and solution of fluid-flow problems. Despite these advances, many ML approaches in CFD remain strongly dependent on simulation or experimental data. In many cases, they are formulated as surrogate or reduced-order models rather than as direct solution strategies for the governing equations \cite{arnold2022large,mohan2018deep,leask2021modal}. In addition, purely data-driven architectures do not automatically satisfy physical constraints such as conservation laws, boundary conditions, or constitutive relationships unless these are explicitly enforced during training \cite{pal2025solving,sluzalec2026reliable,wang_recent_2024}. These limitations motivated the emergence of physics-informed learning as a viable alternative within scientific ML \cite{eivazi_physics-informed_2022,raissi_physics-informed_2019}. By embedding governing equations directly into the learning process, physics-informed methods offer a pathway toward more physically consistent and data-efficient models for fluid-flow problems.

Physics-Informed Neural Networks (PINNs) are a deep learning–based framework that embeds physical laws, physical constraints, and observation data, as introduced by \cite{raissi_physics-informed_2019}. Since their introduction, PINNs have become well-suited for both forward and inverse problems \cite{cai_physics-informed_2021, eivazi_physics-informed_2022, wang_recent_2024}. They have been successfully applied in conjunction with experimental data using various formulations of the NS equations for incompressible flows \cite{eivazi_physics-informed_2022, jin_nsfnets_2021, raissi_physics-informed_2019}, compressible flows \cite{kumar_robust_2026, mao_physics-informed_2020}, and biomedical flow applications \cite{voorter_improving_2023, yin_simulating_2022}. In addition, the NS equations can be combined with thermodynamic constraints \cite{cai_physics-informed_2021, fowler_physics-informed_2024}, which can be incorporated into the neural network loss function by penalizing deviations from the target physical conditions and appropriately weighting them with the available data \cite{cai_physics-informed_2021}. At present, however, the extent to which PINNs can compete with established CFD solvers remains debated. Eivazi et al. \cite{eivazi_physics-informed_2022} demonstrated that PINNs can achieve comparable results under certain conditions, whereas other researchers have questioned whether PINNs can reliably attain the accuracy and robustness of conventional CFD methods \cite{sluzalec2026reliable}. This is because current PINN models often suffer from optimization difficulties associated with minimizing a high-dimensional, non-convex loss function, which can lead to slow convergence, poor stability, and inaccurate solutions, particularly for complex flows \cite{cai_physics-informed_2021, ganga_exploring_2024, zhao_comprehensive_2024}. Their performance is also sensitive to network architecture, sampling strategy, loss weighting, and training procedure \cite{cai_physics-informed_2021, mao_physics-informed_2020, raissi_physics-informed_2019}. Furthermore, PINNs generally remain less accurate and less computationally efficient than high-order CFD methods \cite{cai_physics-informed_2021, karniadakis_spectralhp_2013, wang_recent_2024}, especially for problems involving turbulence, sharp gradients, or multi-scale flow physics. As a result, PINNs currently rely heavily on experimental or simulated data and are primarily used as surrogate models or complementary tools to fill data gaps rather than as full replacements for conventional CFD approaches.

This reliance on experimental or simulated data also highlights a broader need for data-free ML frameworks in fluid mechanics. Such approaches are particularly attractive for problems where high-quality datasets are sparse, expensive to generate, or unavailable across the full range of flow conditions of interest. In theory, a physics-based learning model should be able to solve the governing equations directly while preserving physical consistency and reducing dependence on precomputed data \cite{pal2025solving,sluzalec2026reliable}. However, several bottlenecks in current PINN formulations still limit this capability. In standard PINNs, the residuals of the governing equations are enforced in a point-wise manner through automatic differentiation (AD) \parencite{baydin2018automatic} at collocation points, without explicitly incorporating neighbour-to-neighbour coupling, control-volume flux balances, or stencil-based transport interactions that are central to classical discretization methods. This weakens the direct representation of spatial coupling, particularly in advection-dominated flows, sharp shear layers, and strongly nonlinear regions.

In addition, the representation of first- and second-order derivative terms depends strongly on the smoothness and spectral approximation properties of the selected activation functions, which directly affect the accuracy of pressure gradients, viscous diffusion, and nonlinear convective transport. \textcite{raissi_physics-informed_2019} noted that improved activation functions may help mitigate vanishing-gradient effects and enhance the approximation of higher-order differential operators in deep PINN architectures. Consistent with this, \textcite{jin_nsfnets_2021} showed that commonly used activation functions such as ReLU lack continuous second-order differentiability and are therefore unsuitable for standard NS PINNs formulations, whereas smoother functions such as \textit{tanh} provide improved gradient behavior. Similarly, \textcite{eivazi_physics-informed_2022} reported favorable convergence characteristics with \textit{tanh} for incompressible flow simulations while also recommending the evaluation of alternative activation functions. Furthermore, commonly adopted PINN training strategies, including static loss weighting, random collocation sampling, and gradient-based optimization, can lead to imbalanced residual minimization, spectral bias, and slow or unstable convergence, especially for turbulent or multi-scale flows \parencite{krishnapriyan2021characterizing,wang2021understanding}. Consequently, despite their physics-informed formulation, current PINNs approach remains limited in delivering robust and truly data-independent predictive capability for complex fluid-mechanics applications \cite{ pal2025solving, sluzalec2026reliable}.

In this study, we develop a data-free, physics-informed machine-learning framework for approximating steady incompressible NS solutions across multiple 2D- and 3D- flow configurations and a wide range of Reynolds numbers. The framework learns directly from the governing equations and boundary conditions without using precomputed numerical or experimental data during training. Discretization-inspired differential operators and several training strategies are incorporated to improve the accuracy and robustness of the solutions. The effects of activation functions on learning the governing flow physics are also investigated. Finally, the predicted solutions are compared with direct numerical simulation (DNS) results used exclusively for evaluation, allowing us to assess whether the framework recovers physically meaningful flow behavior without data-based supervision.

\section{Methodology}
In this section, we present the methodological foundation of \textit{NeuralFlowNet}, a specialized PINNs framework for solving the NS equations in 2D- and 3D- domains. \textit{NeuralFlowNet} solves fluid-flow problems directly from the governing equations in both 2D and 3D spaces without requiring measured or simulation-derived training data. The framework employs a multi-layer perceptron (MLP) to map spatial coordinates to the corresponding velocity components and pressure, while the residuals of the governing equations are minimized during training to ensure physical consistency of the predicted solution \cite{raissi_physics-informed_2019,cai_physics-informed_2021}.
\subsection{Multi-Layer Perceptron (MLP)}

Deep neural networks are universal approximators capable of representing highly nonlinear mappings through the composition of multiple layers of transformations. In this work, we adopt a MLP, a fully connected feed-forward neural network, to approximate the target solution field. For an input vector $\mathbf{x}$, the output of layer $l$ is given by
\begin{equation}{\label{eq1}}
\mathbf{z}^{(l)} = g^{(l)}\left(\mathbf{W}^{(l)}\mathbf{z}^{(l-1)} + \mathbf{b}^{(l)}\right),
\end{equation}
where $\mathbf{z}^{(0)}=\mathbf{x}$, and $\mathbf{W}^{(l)}$, $\mathbf{b}^{(l)}$, and $g^{(l)}$ represent the trainable weight matrix, bias vector, and activation function of layer $l$, respectively. The network therefore defines a parametric mapping $f_{\theta}:\mathcal{X}\rightarrow\mathcal{Y}$, with parameters $\theta$ ($\mathbf{W}^{(l)}$, and $\mathbf{b}^{(l)}$) determined through minimization of a prescribed loss function. In the present physics-informed framework, the activation function plays a central role because the governing equations involve both first- and second-order derivatives, which must be represented accurately by the network. Since the ability to represent these derivative terms is directly influenced by the activation function, selecting an appropriate nonlinear function is critical for solving the NS equations \cite{raissi_physics-informed_2019}. Based on our sensitivity analysis, the conventional $\tanh$ activation showed the poorest performance, whereas SiLU \cite{elfwing2018sigmoid} and $x\tanh(x)$ provided improved accuracy and stability in representing both first- and second-order terms. Therefore, SiLU  was adopted in the present work.

\subsection{Physics-Informed Neural Networks (PINNs)}

In classical physics-informed neural networks (PINNs), the spatial and temporal coordinates, $(\mathbf{x},t)$, are given as inputs to a MLP, while the flow variables (velocity field, and pressure) are treated as outputs. For incompressible flow problems, the network defines a mapping
\begin{equation}{\label{eq2}}
\hat{\mathbf{q}}(\mathbf{x},t) = f_{\theta}(\mathbf{x},t),
\end{equation}
where $\hat{\mathbf{q}} = [\hat{u}, \hat{v}, \hat{w}, \hat{p}]$ denotes the predicted velocity and pressure fields, and $\theta$ represents the trainable network parameters. The governing equations consist of the continuity equation (divergence),
\begin{equation}\label{eq3}
\nabla \cdot \mathbf{u} = 0,
\end{equation}
and the momentum equations,
\begin{equation}
\frac{\partial \mathbf{u}}{\partial t}
+
(\mathbf{u}\cdot\nabla)\mathbf{u}
=
-\frac{1}{\rho}\nabla p
+
\nu \nabla^2 \mathbf{u},
\end{equation}{\label{eq4}}
where $\mathbf{u}$ is the velocity vector, $p$ is pressure, $\rho$ is fluid density, and $\nu$ is the kinematic viscosity. In PINNs implementation, these equations are written in residual form using the network predictions during training. Because the network is composed of differentiable operations, the AD evaluates the required derivatives through repeated application of the chain rule, including $\partial \hat{u}_i/\partial x_j$, $\partial^2 \hat{u}_i/\partial x_j^2$, $\partial \hat{u}_i/\partial t$, and $\partial \hat{p}/\partial x_i$ \cite{baydin2018automatic, paszke2017automatic}. These derivatives are then used to construct the total loss function based on the residual of the governing equations, boundary conditions, and initial condition,
\begin{equation}{\label{eq5}}
\mathcal{L}_{\text{PINN}}
=
\lambda_r\mathcal{L}_r
+\lambda_b\mathcal{L}_b
+\lambda_i\mathcal{L}_i
+\lambda_d\mathcal{L}_d,
\end{equation}
where $\mathcal{L}_r$, $\mathcal{L}_b$, $\mathcal{L}_i$, and $\mathcal{L}_d$ represent the partial differential equation (PDE) residual, boundary-condition residual, initial-condition residual, and optional data losses, respectively, and $\lambda_r$, $\lambda_b$, $\lambda_i$, and $\lambda_d$ are their corresponding weighting coefficients. Although this formulation is elegant, classical PINNs often struggle for advection-dominated and turbulent flows, where nonlinear transport and diffusion must both be represented accurately \cite{cengizci2026physics, lu2026gradient,pal2025solving}.

\subsection{PINNs Adaptation in NeuralFlowNet}

\begin{figure*}[!h]
    \centering
    \includegraphics[width=1\textwidth,height=0.6\textheight,keepaspectratio]{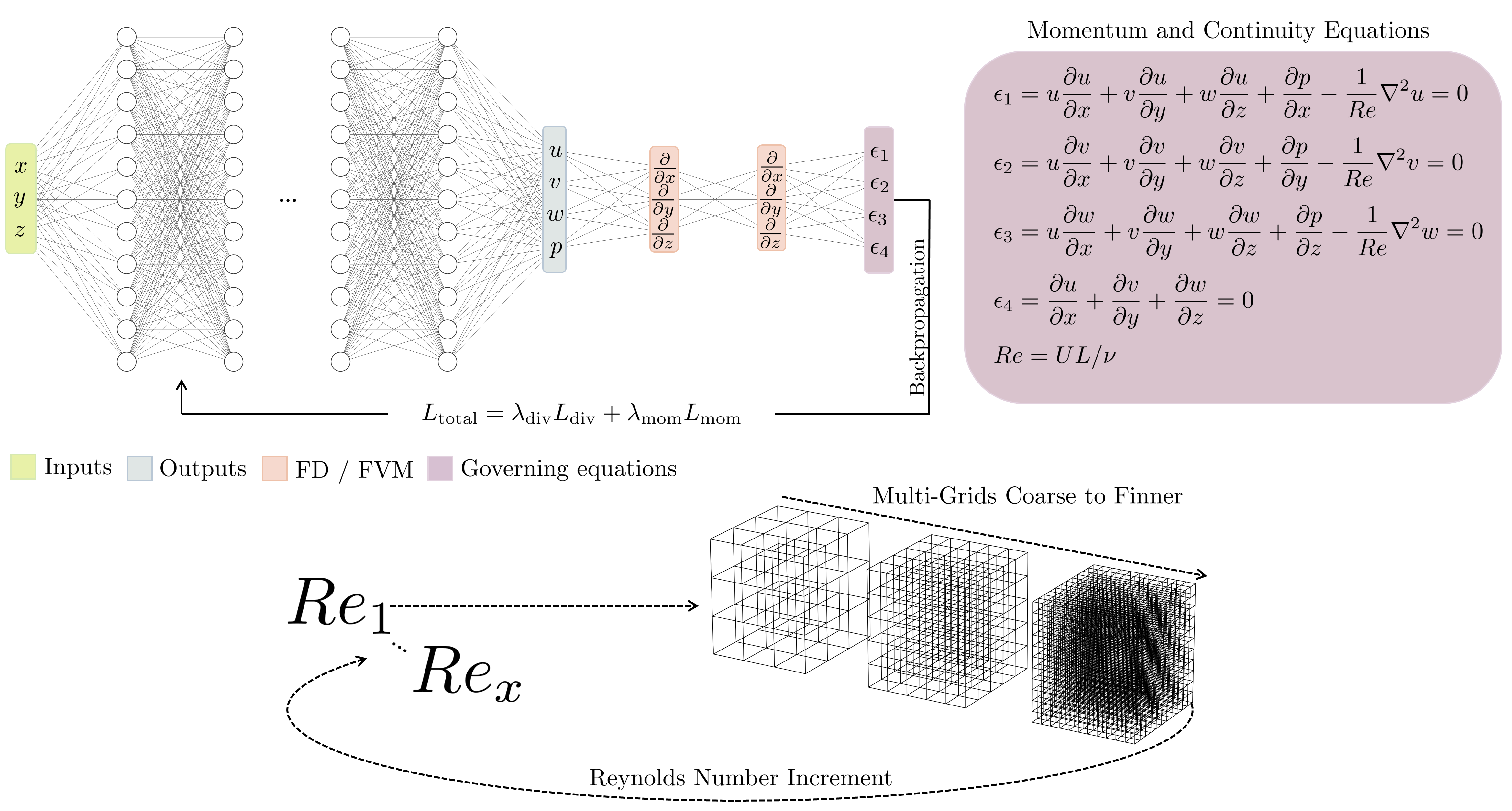}
    \caption{Schematic of the \textit{NeuralFlowNet} framework for solving the 3D NS equations, illustrating the general set-up and the iterative progression from coarse to fine grids across different Reynolds numbers.}
    \label{fig1}
\end{figure*}

\subsubsection{MLP Prediction}
The MLP predicts flow variables and pressure over the computational domain, namely $u$, $v$, and $p$ for 2D-, and $u$, $v$, $w$, and $p$ for 3D- cases. Since these variables are generated directly by the network, at early iterations the simulations are not guaranteed to satisfy either the prescribed boundary conditions or the governing equations. Therefore, the network must be constrained to satisfy both the boundary conditions and the governing physics during training.

\subsubsection{Boundary Condition Enforcement}
To enforce the boundary conditions exactly, they are imposed by explicitly overwriting the predicted boundary values at every iteration \cite{pal2025solving}, also known as a hard boundary condition. Depending on the governing variable and the physical configuration, this includes Dirichlet boundary conditions, which prescribe the value of velocity or pressure at the boundary, and Neumann boundary conditions \parencite{pal2025solving}, which prescribe the corresponding gradient \cite{berrone2023enforcing,gladstone2025fo,pal2025solving}. This hard-boundary treatment eliminates the need for an additional learnable boundary-loss weight and ensures exact satisfaction of the boundary constraints through the entire domain \cite{pal2025solving}. However, exact boundary enforcement alone does not guarantee that the interior solution is physically consistent. The governing operators must also be evaluated in a manner that captures the interaction between neighboring cells, which is essential in fluid-flow problems governed by transport processes.
\subsubsection{Discrete Operator for First and Second Derivative}
From our experiments, we conclude that the key limitation of classical PINNs, is that AD evaluates the governing equations pointwise through the chain rule without explicitly accounting for neighboring-point interactions. For incompressible flow and NS in non-dimensional form, the governing dynamics are controlled by the combined action of different components, nonlinear advection $(\mathbf{u}\cdot\nabla)\mathbf{u}$, pressure-gradient $\nabla p$, the viscous diffusion $(1/Re)\nabla^2\mathbf{u}$, and the divergence-free constraint $\nabla\cdot\mathbf{u}=0$, which propagate the errors of the approximations in intricate manners during training.

To capture better each one of those components \textit{NeuralFlowNet} adopts a numerical and CFD-inspired approximation, by using finite-difference (2D)  and finite-volume (3D), so that each component is evaluated using neighboring-cell information rather than a pointwise estimation through AD. The method used to approximate the derivatives is very important to ensure convergence in fluid dynamic \cite{godunov1959finite}. It is well known in numerical methods that the direction of the flow and the order of the derivative must be considered in the moment we approximate a derivative, which is something that is not considered by AD. Therefore,  we consider that the use of finite-difference and finite-volume is a more rigorous manner to proceed. The details of selecting those methods are explained in a 2D case, however the extension to three dimensions is straightforward and follows the same framework with the inclusion of the additional velocity component and corresponding terms.

As it is known in the specialized literature \cite{gerhart2021munson,ghia1982high,xiao2012simulating,xiao2020flows}, the consideration of approximations (finite difference or finite volume) in each component of the NS equation introduces errors that must be addressed. Therefore, we present each consideration below.\\

\noindent\textbf{\textit{Advection}}\\
Among the NS equation, the advection, $(\mathbf{u}\cdot\nabla)\mathbf{u}$, is the most numerically sensitive component because it governs nonlinear momentum transport and is strongly affected by the trade-off between numerical stability and numerical diffusion \cite{ferziger2002computational}. For this reason, a purely upwind discretization is more stable during the early stages of training \cite{koren1990upwind}, because the predicted solution is still far from the physical state. However, it introduces artificial numerical diffusion that can smear important flow structures \cite{ferziger2002computational}. In contrast, a purely central discretization is less diffusive and better preserves gradients, but it is more prone to oscillations and instability \cite{brooks_streamline_1982,ferziger2002computational}. To balance these competing effects, \textit{NeuralFlowNet} evaluates the advection term using a blended formulation of both discretizations.
\begin{equation}{\label{eq6}}
\mathcal{A} = (1-\beta)\mathcal{A}_{\mathrm{upwind}} + \beta \mathcal{A}_{\mathrm{central}},
\end{equation}
where $\mathcal{A}$ is the advection component evaluated in NS, $\mathcal{A}_{\mathrm{upwind}}$ and $\mathcal{A}_{\mathrm{central}}$ are the upwind and central contributions, respectively, and $\beta$ is a blending factor inspired by \parencite{greenshields_openfoam_2021}. In practice, $\beta$ is bounded within $0 \leq \beta \leq 1$, allowing the scheme to vary from more stable to sharper treatments without collapsing to either extreme \cite{ferziger2002computational}. Moreover, $\beta$ is updated adaptively during training based on PDE-balance diagnostics. In particular, to tailor the advection treatment to the evolving flow field, $\beta$ is adjusted using several dimensionless diagnostic ratios.\\
The first diagnostic is the diffusion-dominance ratio \cite{ferziger2002computational,van1977towards},
\begin{equation}
\label{eq9}
\small
\begin{aligned}
r_\nu
= \frac{1}{2}
\Bigg[
&\frac{\|(1/Re)\nabla^2 u\|}
{\|(\mathbf{u}\cdot\nabla)u\| + \|\partial p/\partial x\| + \epsilon} \\
&+
\frac{\|(1/Re)\nabla^2 v\|}
{\|(\mathbf{u}\cdot\nabla)v\| + \|\partial p/\partial y\| + \epsilon}
\Bigg].
\end{aligned}
\end{equation}
where $\epsilon$ is a small positive constant introduced to avoid division by zero. This ratio measures whether diffusion is becoming too dominant relative to the combined effects of advection and pressure. From our experimentation, values of $r_\nu > 0.80$ indicate an over-diffused state, whereas values of $r_\nu < 0.40$ indicate that diffusion has weakened sufficiently to permit a sharper treatment of the advection  term.

The second diagnostic is the divergence-to-momentum ratio \cite{ferziger2002computational,greenshields_openfoam_2021},
\begin{equation}\label{eq10}
r_{\mathrm{divmom}}
=
\frac{\|\nabla\cdot\mathbf{u}\|}
     {\|R_{\mathrm{mom}}\| + \epsilon},
\end{equation}
where $\|R_{\mathrm{mom}}\|$ is the root-mean-square momentum residual obtained from the assembled momentum equations. This ratio measures whether divergence errors are becoming too large relative to the momentum residual; values of $r_{\mathrm{divmom}} > 0.70$ indicate a divergence-limited state, values below $0.45$ are treated as acceptable for sharper advection treatment, and values below $0.25$ represent a strongly divergence-controlled state.

The third diagnostic is the advection-strength ratio \cite{bayareh2021artificial,ferziger2002computational},
\begin{equation}\label{eq11}
r_{\mathrm{adv}}
=
\frac{\|(\mathbf{u}\cdot\nabla)u\| + \|(\mathbf{u}\cdot\nabla)v\|}
     {\|(1/Re)\nabla^2 u\| + \|(1/Re)\nabla^2 v\| + \epsilon}.
\end{equation}
which measures the relative importance of advection compared with diffusion. This ratio is used to determine whether the solution has entered an advection-dominated regime. In the present controller, values of $r_{\mathrm{adv}} > 1.20$ mark the onset of advection-dominant behavior, whereas values of $r_{\mathrm{adv}} > 2.00$ indicate a strongly advection-driven state.

Finally, the numerical-diffusion ratio is defined as
\begin{equation}\label{eq12}
r_{\mathrm{num}} = \frac{0.5(\nu_{\mathrm{num},x}+\nu_{\mathrm{num},y})}{\nu+\epsilon},
\end{equation}
where $\nu_{\mathrm{num},x}$ and $\nu_{\mathrm{num},y}$ denote the estimated artificial numerical viscosities associated with the discretization in the two coordinate directions, and $\nu$ is the physical viscosity of the flow. In the present implementation, these directional numerical viscosities are estimated as
\begin{equation}\label{eq13}
\nu_{\mathrm{num},x}=\frac{1}{2}(1-\beta)\,|u|\,\Delta x\,(1-c_x),
\end{equation}
\begin{equation}\label{eq14}
\nu_{\mathrm{num},y}=\frac{1}{2}(1-\beta)\,|v|\,\Delta y\,(1-c_y).
\end{equation}
Here, $|u|$ and $|v|$ are the absolute values of the local velocity components, while $\Delta x$ and $\Delta y$ are the local grid spacings in the streamwise and transverse directions, respectively. The formulation follows the modified-equation interpretation of first-order upwind discretization, in which the leading truncation error behaves as numerical diffusion proportional to the local transport speed and grid spacing \cite{ferziger2002computational,greenshields_openfoam_2021,kyurkchiev2015sigmoid,menon1996characterization}. The factor $(1-\beta)$ represents the upwind contribution of the blended convection operator, while the bounded correction factors $c_x,c_y\in[0,1]$ control the retained numerical viscosity in each coordinate direction. In the present implementation, these control parameters are prescribed through an adaptive target value and selected empirically through preliminary numerical experiments and iterative controller calibration. Thus, the estimated numerical viscosity increases when the upwind contribution becomes stronger, when the local velocity magnitude increases, and when the control parameter is reduced. The ratio $r_{\mathrm{num}}$ therefore quantifies the relative importance of artificial numerical diffusion compared with the physical viscosity. In the present controller, values of $r_{\mathrm{num}} > 0.25$ indicate that numerical viscosity has become significant, whereas $r_{\mathrm{num}} = 0.10$ is used as the nominal target level.

The thresholds for these four ratios were selected empirically through preliminary numerical experiments and iterative controller calibration rather than being derived from strict analytical bounds. Different candidate ranges were tested to identify values that could reliably distinguish over-diffused, divergence-limited, advection-dominated, and numerically diffusion-dominated solution states while maintaining stable adaptive updates during training. The final thresholds were retained because they yielded smooth controller behavior, reduced erratic switching between numerical regimes, and produced consistent convergence across the benchmark cases considered in this study.

These diagnostic ratios are then used to update the blending factor $\beta$ during training. However, given the rapid oscillation those ratios could face during training, a smooth evolution and transition was implemented. Each one of the ratios is updated just partially in each iteration through this equation
\begin{equation}\label{eq15A}
r^{\mathrm{updated}}
=
(1-\alpha)r^{\mathrm{previous}}
+
\alpha r^{\mathrm{instantaneous}},
\end{equation}
where $\alpha=0.08$ is the prescribed updating coefficient. Therefore, each updated diagnostic ratio retains 92\% of its previous value and incorporates 8\% of the current instantaneous diagnostic ratio. The resulting ratios are then standardized relative to their corresponding controller reference levels, namely $r_{\mathrm{num,target}}=0.10$, $r_{\mathrm{adv,ready}}=1.20$, and $r_{\mathrm{div,mid}}=0.45$, as
\begin{equation}\label{eq15}
z_{\mathrm{num}}=\frac{{r_{\mathrm{num}}^{updated}}-r_{\mathrm{num,target}}}{0.08},
\end{equation}
\begin{equation}\label{eq16}
z_{\mathrm{adv}}=\frac{{r_{\mathrm{adv}}^{updated}}-r_{\mathrm{adv,ready}}}{0.30},
\end{equation}
and
\begin{equation}\label{eq17}
z_{\mathrm{div}}=\frac{{r_{\mathrm{divmom}}^{updated}}-r_{\mathrm{div,mid}}}{0.15}.
\end{equation}
Here, $r_{\mathrm{num,target}}$, $r_{\mathrm{adv,ready}}$, and $r_{\mathrm{div,mid}}$ denote the reference levels about which the adaptive controller is centered, while the denominators define the common ranges of the instantaneous ratio distribution. Those values were determined through trial and error by evaluating the correct standardization.   

The smoothing (equation \ref{eq15A}) is introduced to reduce sensitivity to short-term fluctuations in the raw ratios. The standardized (equations \ref{eq15}-\ref{eq17}) quantities are then passed through sigmoid functions $\sigma(\cdot)$, which were selected because they provide a bounded, and monotonic mapping from the diagnostic variables to the controller response \cite{godunov1959finite,kyurkchiev2015sigmoid,menon1996characterization}. This avoids abrupt switching between numerical regimes, limits excessively large updates, and retains sensitivity near the transition region where changes in the advection treatment are most important. The target value of the blending factor is then constructed as
\begin{equation}
\label{eq18}
\begin{aligned}
\beta_{\mathrm{target}} \;=\;& a_0
+ a_1\,\sigma(z_{\mathrm{num}})
+ a_2\,\sigma(z_{\mathrm{adv}}) \\
& - a_3\,\sigma(z_{\mathrm{div}})
+ a_4\,t,
\end{aligned}
\end{equation}
with $(a_0,a_1,a_2,a_3,a_4)=(0.05,\,0.55,\,0.35,\,0.45,\,0.15)$. Those values were obtained by solving the logistic regression problem after defining some critical conditions that could be faced during training. Here $a_0$ provides a small baseline that prevents $\beta_{\mathrm{target}}$ from collapsing to zero in the absence of strong diagnostic signals. The positive weights $a_1$ and $a_2$ increase $\beta_{\mathrm{target}}$ when the corresponding indicators suggest that numerical diffusion is excessive and/or that an advection-dominant regime has been reached, thereby promoting a sharper (more central) transport treatment. In contrast, the divergence-related contribution is assigned a negative sign with coefficient $a_3$ so that elevated divergence (relative to momentum balance) reduces $\beta_{\mathrm{target}}$, biasing the scheme toward a more dissipative upwind contribution when additional robustness is required. Finally, the term $a_4\,t$ introduces a mild progression with training, encouraging $\beta_{\mathrm{target}}$ to increase gradually as optimization proceeds, while remaining moderated by the diagnostic terms.
where $t$ is defined by
\begin{equation}\label{eq8}
t = \frac{ep}{\max(E-1,1)},
\end{equation}
where $ep$ is the current epoch and $E$ is the total number of epochs in the stage. 

The resulting $\beta_{target}$ value is then further modified according to the detected diagnostic regime $\mathcal{R}$ defined as
\begin{equation}\label{eq:regime_conditions}
\mathcal{R} =
\begin{cases}
\text{divergence-limited}, 
& r_{\mathrm{divmom}} > 0.70,\\[5pt]

\text{over-diffused}, 
& r_{\nu} > 0.80,\quad
  r_{\mathrm{adv}} < r_{\mathrm{adv,ready}},\\[5pt]

\text{advection-ready}, 
& \begin{aligned}[t]
  & r_{\mathrm{adv}} > r_{\mathrm{adv,ready}},\quad
    r_{\nu} < 0.40,\\
  & r_{\mathrm{divmom}} < r_{\mathrm{div,mid}}.
\end{aligned}
\end{cases}
\end{equation}
with the following threshold modifications 
\begin{equation}\label{eq19}
\beta_{\mathrm{target}}=
\begin{cases}
\min\!\left(\beta_{\mathrm{target}},\,0.55\right), & \text{divergence-limited},\\[6pt]
\min\!\left(\beta_{\mathrm{target}}+0.10,\,0.75\right), & \text{over-diffused},\\[6pt]
\min\!\left(\beta_{\mathrm{target}}+0.10,\,1.00\right), & \text{advection-ready}.
\end{cases}
\end{equation}
If none of these regime conditions is satisfied, $\beta_{\mathrm{target}}$ remains equal to $\beta_{\mathrm{target}}$. The numerical bounds we used in this regime-based modification were selected through preliminary numerical experiments and iterative controller calibration rather than from strict analytical bounds. These values were retained because they provided stable transitions between numerical regimes, avoided overly aggressive sharpening of the advection term, and produced smooth and consistent convergence across the benchmark cases considered in this study. Accordingly, $\beta_{\mathrm{target}}$ increases when numerical diffusion becomes too strong, and advection becomes more important, and decreases when divergence errors remain comparatively large. Thus, when the evolving solution is sufficiently stable, divergence is controlled, advection strengthens, and numerical diffusion remains limited, $\beta_{\mathrm{target}}$ is increased so that the advection term is treated more sharply. Conversely, when the solution remains numerically fragile, $\beta_{\mathrm{target}}$ is reduced so that the formulation remains more upwind-dominant and stable. 

After these regime-based corrections are applied, the final blending factor is updated through the same updating methods described in eq. \ref{eq15A} \cite{greenshields_openfoam_2021}.

\begin{equation}{\label{eq20}}
\beta^{updated}
=
(1-\alpha_{\beta})\beta^{previous}
+\alpha_{\beta}\beta_{\mathrm{target}},
\end{equation}

where $\beta^{\mathrm{previous}}$ denotes the blending factor from the previous controller update, and $\beta^{\mathrm{updated}}$ denotes the new relaxed blending factor used in the advection operator in equation~\ref{eq6}. At the beginning of training, $\beta$ is initialized with 0; after this initialization, the adaptive controller updates $\beta$ according to equation~\ref{eq20}. The parameter $\alpha$ is the same updating coefficient described in equation \ref{eq15A}. The $\beta^{updated}$ value is then subjected to a final clamp to ensure that $\beta$ remains within its admissible range (i.e. $0 \leq \beta \leq 1$). This updating process prevents abrupt switching in the advection treatment and allows $\beta$ to evolve gradually with the numerical state of the solution. Because the upwind contribution also introduces artificial numerical diffusion, this effect is accounted for separately in the following section.\\

\noindent\textbf{\textit{Diffusion}}\\
The diffusion, represented by $(1/Re)\nabla^2\mathbf{u}$, describes physical viscous momentum transport and is evaluated using central differencing. Unlike the advection term, the second-order viscous operator is symmetric and is therefore naturally suited to a central approximation. However, the effective diffusive behavior of the solution is not governed solely by the physical viscosity, because the upwind contribution used in the advection treatment introduces an additional artificial diffusion-like effect \cite{brooks_streamline_1982,jameson1993artificial}. If left uncorrected, this numerical diffusion can lead to excessive smoothing of the predicted flow field.

To account for this behavior, \textit{NeuralFlowNet} estimates the artificial numerical viscosity introduced by the discretization and incorporates an effective-viscosity correction \cite{ferziger2002computational},
\begin{equation}\label{eq21}
\nu_{\mathrm{eff},i}
=
\max\left(\nu-\nu_{\mathrm{num},i},\,\nu_{\min}\right),
\qquad i \in \{x,y\},
\end{equation}
where $\nu_{\mathrm{num},i}$ denotes the directional numerical-viscosity estimate defined in equations~\ref{eq13} and \ref{eq14}, and $\nu_{\min}$ is a lower bound on the effective viscosity retained for numerical stability. Here, $\nu_{\mathrm{eff},i}$ should not be interpreted as a modification of the physical fluid viscosity. The physical viscosity remains $\nu = 1/Re$, as prescribed by the NS equations. Instead, this correction is introduced because the upwind-biased treatment of the advection term adds artificial numerical diffusion \cite{godunov1959finite,greenshields_openfoam_2021,ferziger2002computational}, which is not part of the original governing equations and can over-smooth the predicted solution if left unaccounted for. Accordingly, the diffusion operator is adjusted so that part of this added numerical smoothing is compensated within the discrete momentum balance, while still preserving a minimum level of viscosity required for stable training. In the present implementation, this lower bound was taken as a small fraction of the physical viscosity, approximately 1\% - 3\% of $\nu$. Thus, the diffusion treatment in \textit{NeuralFlowNet} reflects both the physical viscous transport prescribed by the NS equations and the additional numerical diffusion introduced by the adaptive advection discretization \cite{ferziger2002computational,greenshields_openfoam_2021}.\\

\noindent\textbf{\textit{Pressure-Gradient}}\\
The pressure field and its spatial gradients are evaluated using central differencing. This choice is adopted because the pressure-gradient term represents a symmetric force contribution in the momentum equations rather than a transported quantity. Therefore, a centered approximation avoids the directional bias and artificial diffusion associated with upwind schemes \cite{ferziger2002computational,van1977towards}. These pressure-gradient terms are then inserted directly into the momentum residuals so that the learned pressure field contributes explicitly to the force balance.\\

\noindent\textbf{\textit{Projection Treatment}}\\
To further improve satisfaction of the incompressibility constraint, \textit{NeuralFlowNet} applies a projection-based velocity correction to the provisional velocity field, following the classical projection approach for incompressible-flow solvers \parencite{chorin1997numerical,guermond1998stability,ferziger2002computational}. In this step, the provisional velocity $\mathbf{u}^*$ predicted by the network is corrected as
\begin{equation}\label{eq22}
\mathbf{u}^{\mathrm{corr}} = \mathbf{u}^* - \nabla \phi,
\end{equation}
where $\phi$ is a scalar correction potential. The correction potential is obtained by solving the Poisson equation
\begin{equation}\label{eq23}
\nabla^2 \phi = \nabla \cdot \mathbf{u}^*,  
\end{equation}
so that the corrected velocity field is driven closer to the divergence-free condition. This can be seen by taking the divergence of equation~\ref{eq22}, which gives
\begin{equation}\label{eq24}
\nabla \cdot \mathbf{u}^{\mathrm{corr}}
=
\nabla \cdot \mathbf{u}^*
-
\nabla^2 \phi .
\end{equation}
Therefore, when $\phi$ satisfies equation~\ref{eq23}, the divergence contribution from $\mathbf{u}^*$ is reduced, and the corrected velocity field approaches $\nabla \cdot \mathbf{u}^{\mathrm{corr}}\approx 0$.
In practice, the projection equation is solved iteratively, so the correction reduces rather than completely eliminates the divergence error.

In the present implementation, the projection correction is not activated at the beginning of each multigrid training stage. Instead, projection is disabled during the initial warm-up period, defined as the first 30\% of the training epochs in the current stage. This warm-up period allows the velocity and pressure fields predicted by the neural network to develop before an additional divergence-reduction correction is imposed. After the warm-up period, the projection correction is applied adaptively based on the divergence-to-momentum diagnostic, $r_{\mathrm{divmom}}$. At a given training epoch, the network first predicts a provisional velocity field. This provisional velocity field is not guaranteed to be exactly divergence-free. Therefore, the divergence of the provisional velocity field is computed and used as the source term for a correction-potential Poisson equation. The purpose of this Poisson problem is not to solve for the physical pressure directly, but to obtain a scalar correction potential whose gradient can be used to reduce the divergence of the provisional velocity field.

The Poisson equation is solved approximately using Jacobi relaxation \cite{roy2015parallel,zhao2026fourth}. In this implementation, the correction potential is initialized to zero at the beginning of each projection step. The Jacobi method then updates the correction potential repeatedly using the divergence of the provisional velocity field as the source term. Each complete Jacobi relaxation sweep over the computational grid is referred to as one Poisson iteration. Therefore, one Poisson iteration is one Jacobi relaxation step used to improve the approximate solution of the correction-potential Poisson equation. The Poisson problem is not solved to full convergence at every training epoch. Instead, only a prescribed number of Poisson-solver iterations is performed, and this number defines the projection strength. A larger number of Poisson iterations gives a stronger projection correction because the correction potential is updated more times, producing a stronger divergence-reducing velocity correction. A smaller number of iterations gives a weaker correction, while zero iterations means that the projection correction is disabled.

The number of Poisson-solver iterations is selected adaptively from $r_{\mathrm{divmom}}$. When $r_{\mathrm{divmom}}>0.70$, the solution is classified as divergence-limited, and the strongest projection correction is applied using the maximum number of Poisson-solver iterations. In the present implementation, this corresponds to 25 Poisson-solver iterations. For moderate divergence, $0.45<r_{\mathrm{divmom}}\leq0.70$, an intermediate correction is applied using approximately 11 Poisson-solver iterations. For mild divergence, $0.25<r_{\mathrm{divmom}}\leq0.45$, a weaker correction is applied using approximately 4 Poisson-solver iterations. When $r_{\mathrm{divmom}}\leq0.25$, divergence is considered sufficiently controlled, and the projection correction is disabled by using no Poisson iterations.

After the prescribed number of Poisson-solver iterations is completed, the gradient of the correction potential is computed and subtracted from the provisional velocity field. This produces a corrected velocity field with reduced divergence. The corrected velocity field is then used in the loss evaluation and residual calculation during training. In this way, the projection step acts as a numerical divergence-control operation embedded inside the training loop, rather than as a separate fully converged pressure solver. The threshold values 0.70, 0.45, and 0.25 are empirical controller thresholds selected from preliminary calibration experiments to distinguish divergence-limited, moderate-divergence, mild-divergence, and divergence-controlled regimes during training. Therefore, the warm-up length is determined by the number of epochs assigned to each multigrid training stage, while the projection strength is determined dynamically from the evolving flow diagnostics.

\subsubsection{Residual calculation}

\noindent\textbf{\textit{Momentum Residual}}\\
Having treated the advection component, the pressure-gradient component, and the viscous diffusion component, separately according to their numerical behavior, these contributions are then assembled into the discrete momentum residual. In compact form, this residual may be written as
\begin{equation}\label{eq24}
\mathbf{R}_{\mathrm{mom}} = (\mathbf{u}\cdot\nabla)\mathbf{u} + \nabla p - \frac{1}{Re}\nabla^2\mathbf{u},
\end{equation}
where $\mathbf{R}_{\mathrm{mom}}$ denotes the momentum residual vector. Although this expression is written in vector form, the implementation evaluates the residual component-wise over the computational domain.\\ 

\noindent\textbf{\textit{Divergence Residual}}\\
The continuity equation is written as the divergence-free condition (equation \ref{eq3}) and the corresponding divergence residual is defined as
\begin{equation}\label{eq25}
\mathbf{R}_{\mathrm{div}} = \nabla \cdot \mathbf{u}.
\end{equation}
In the implementation, this quantity is evaluated using central differencing over the interior of the computational domain. 

\subsubsection{Loss and Optimization}
Once the discrete momentum and divergence residuals have been calculated, the corresponding loss terms are defined as mean-squared residual measures over the computational domain. For the 2D formulation presented here, the momentum loss is written as
\begin{equation}\label{eq26}
L_{\mathrm{mom}} = \lambda_u \,\mathrm{mean}(\mathbf{R_u^2}) + \lambda_v \,\mathrm{mean}(\mathbf{R_v^2}),
\end{equation}
where $\mathbf{R}_u$ and $\mathbf{R}_v$ denote the residuals of the $x$- and $y$-momentum equations, respectively. Additional weighting factors, $\lambda_u$ and $\lambda_v$, are applied to the momentum residuals in order to balance the relative contributions of the two momentum equations within the momentum loss. In the present implementation, $\lambda_u$ and $\lambda_v$ were assigned equal weights (i.e. $\lambda_u$ \& $\lambda_v$ = 1.0).

Similarly, the divergence loss is defined as
\begin{equation}\label{eq27}
L_{\mathrm{div}} = \mathrm{mean}(\mathbf{R_{\mathrm{div}}^2}),
\end{equation}
where $\mathbf{R_{\mathrm{div}}}$ denotes the residual of the continuity equation. The total loss is then written as
\begin{equation}\label{eq28}
L_{\mathrm{total}} = \lambda_{\mathrm{div}} L_{\mathrm{div}} + \lambda_{\mathrm{mom}} L_{\mathrm{mom}},
\end{equation}
where $\lambda_{\mathrm{div}}$ and $\lambda_{\mathrm{mom}}$ are epoch-dependent weights used to control the relative contribution of continuity and momentum during training. The equations used by \textit{NeuralFlowNet} were the result of experimentation to balance the diversion and momentum losses adequately. These weights are prescribed through the normalized epoch coordinate in equation \ref{eq8} with
\begin{equation}\label{eq29}
\lambda_{\mathrm{mom}} = 0.20 + 0.80t^2,
\qquad
\lambda_{\mathrm{div}} = 20(1-t)^2 + 2.
\end{equation}

These forms were chosen to reflect the physical and numerical roles of the governing equations during training. This treatment is partly motivated by stream-function and vorticity–streamfunction formulations\parencite{ferziger2002computational,kundu2024fluid}, where velocity is defined to inherently satisfy incompressibility and pressure can be removed from the primary variables. In addition, velocity fields are primarily characterized by spatial structure, whereas pressure may exhibit more pronounced temporal evolution effects \parencite{de2025kan}. In the early epochs, the divergence constraint is emphasized more strongly so that the predicted velocity field is first guided toward incompressibility. As training progresses, the momentum loss is increased so that the solution is refined toward the correct balance among nonlinear transport, pressure forces, and viscous diffusion. Thus, the weighting strategy is physically motivated in its overall structure, while the specific constants were selected through numerical calibration to obtain stable and effective convergence.

The total loss is minimized by backpropagation through the network, such that gradients of $L_{\mathrm{total}}$ with respect to all trainable parameters of the MLP. In the present implementation, the hidden layers employ a SiLU activation function, whose smooth nonlinear response is well suited for stable optimization and gradient-based learning of fluid-flow fields. Training is performed using a staged optimization strategy that combines an initial Adam phase ($\approx$10{,}000--30{,}000 epochs), followed by an L-BFGS refinement ($\approx$1{,}500--3{,}000 epochs), and finally sequential variable-refinement steps \cite{eivazi_physics-informed_2022,liu1989limited}. In this last stage, the pressure field is first held fixed while the velocity components are further optimized. Next, the velocity field is held fixed while the pressure field is refined. Finally, a coupled optimization is performed in which all variables are updated simultaneously. The Adam optimizer is used again during this sequential variable-refinement ($\approx$epochs 500 - 800). Overall, approximately 90\% of the training is carried out using Adam, while the remaining 10\% uses the L-BFGS optimizer. The L-BFGS method terminates automatically when no further improvement in the loss is observed \cite{eivazi_physics-informed_2022}.

Convergence is further enhanced through a multigrid coarse-to-fine transfer strategy \cite{ghia1982high,ferziger2002computational, kolmogorov_local_1941}, in which solutions are progressively refined across grid levels. Rather than focusing on specific grid sizes, the key idea is to follow a consistent trend of increasing resolution, where each stage is moderately finer than the previous one, typically by a factor of about 1.5 to 2. At each level, the solution is iterated until convergence, and the resulting state is then used to initialize the next finer grid. This gradual refinement allows the model to capture large-scale flow features on coarse grids and progressively resolve finer details on denser grids while maintaining stability.

After the model is trained across all grid resolutions, the Reynolds number is increased and the same grid-refinement procedure is repeated. The simulation is first initialized at $Re=100$, where convergence is relatively easier to achieve. From there, the Reynolds number is gradually increased in increments of 100 toward the targeted value, with each new case initialized using the previously converged solution. This stepwise Reynolds-number continuation provides a stable transition to progressively more complex flow regimes. In this work, the cases were systematically tested over a range from Re=100 up to Re=2000, and in some instances up to Re=5000, demonstrating the robustness of the approach across a wide range of flow conditions \cite{rott1990note,sillero2013one}.

\begin{table*}[!b]
\centering
\caption{Summary of benchmark cases, Reynolds numbers, and prediction differences relative to the reference solutions}
\label{tab:benchmark_summary}

\renewcommand{\arraystretch}{1.35}
\setlength{\tabcolsep}{10pt}

\begin{tabularx}{\textwidth}{|c|X|c|c|c|c|}
\hline
\textbf{Case Type} & \textbf{Test} & \textbf{Re} & $\mathbf{NRMSE_u}$ & $\mathbf{NRMSE_v}$ & $\mathbf{NRMSE_p}$ \\
\hline
2D Cases & Lid Driven Cavity & 100  & 0.44\% & 0.69\% & 2.83\%  \\
         & Lid Driven Cavity & 2000 & 1.13\%  & 1.53\%  & 13.97\% \\
         & Channel Flow      & 100  & 0.23\% & 0.15\%  & 0.13\% \\
         & Channel Flow      & 5000 & 1.03\% & 3.11\% & 9.45\% \\
         & Periodic Hill     & 100  & 7.02\%  & 6.50\%  & 17.04\% \\
         & Periodic Hill     & 200  & 5.05\%  & 6.17\%  & 12.98\% \\
         & Periodic Hill     & 400 & 5.27\% & 4.97\% & 11.41\%  \\
\hline
3D Cases & Pipe Flow & 100  & 0.05\% & - & - \\ 
         & Pipe Flow & 3000 & 1.40\% & - & - \\
         & BFS       & 100  & 4.95\% & 6.01\%  &  4.16\%  \\
        & BFS       & 200  & 5.56\% & 7.46\%  & 3.20\% \\
        & BFS       & 400  & 4.39\%  & 8.76\% & 5.46\% \\
\hline
\end{tabularx}
\end{table*}

\subsection{Benchmark Cases and Validation Metrics}

\textit{NeuralFlowNet} is currently formulated for steady-state flow solutions. However, this does not mean it could not be expanded to a dynamic framework. In this research, we focus on demonstrating the capabilities of the approach for solving high Reynolds numbers, and we leave the dynamic considerations for future work. For this reason, its predictive accuracy was evaluated against high-fidelity steady-state reference solutions for both 2D and 3D benchmark cases developed in OpenFOAM (v2312) following configurations reported in the literature \cite{breuer2009flow,ghia1982high,rapp2011flow,kim1980investigation,jovic1994backward, williams1997numerical}. These cases were used to cross-validate the numerical setup and boundary conditions. For each case, the reference field was taken from the fully converged steady-state \texttt{simpleFoam} implementation, where convergence was defined by a pressure tolerance of $10^{-6}$, a velocity tolerance of $10^{-5}$, and the absence of meaningful changes in the solution fields with further Semi-Implicit Method for Pressure Linked Equations (SIMPLE) iterations. For the 2D high-Reynolds-number cases, these solutions are resolved by using the modeling type called "laminar" in the OpenFOAM software. Model performance for the primary variables $u$, $v$, $w$, and $p$ was quantified using both the Normalized Root Mean Square Error ($NRMSE$)\cite{stephen2014improved} over the domain and the pointwise absolute difference. For a generic variable $\phi \in \{u,v,w,p\}$, the relative $NRMSE$ was defined as
\begin{equation}
\mathrm{NRMSE} =
\frac{
\sqrt{
\frac{1}{N}
\sum_{i=1}^{N}
\left(
\phi_{\mathrm{DNS},i}
-
\phi_{\mathrm{NFN},i}
\right)^2
}
}{
\phi_{\mathrm{DNS},\max}
-
\phi_{\mathrm{DNS},\min}
}
\times 100
\end{equation}
and the pointwise absolute difference was defined as
\begin{equation}
E_{\phi,i}^{\mathrm{abs}}=
\left|
\phi_{\mathrm{DNS},i}-\phi_{\mathrm{NFN},i}
\right|,
\end{equation}
where $\phi_{\mathrm{NFN}}$ denotes the \textit{NeuralFlowNet} prediction.

\section{Results}

\begin{figure*}[!b]
    \centering
    \includegraphics[width=18.2cm]{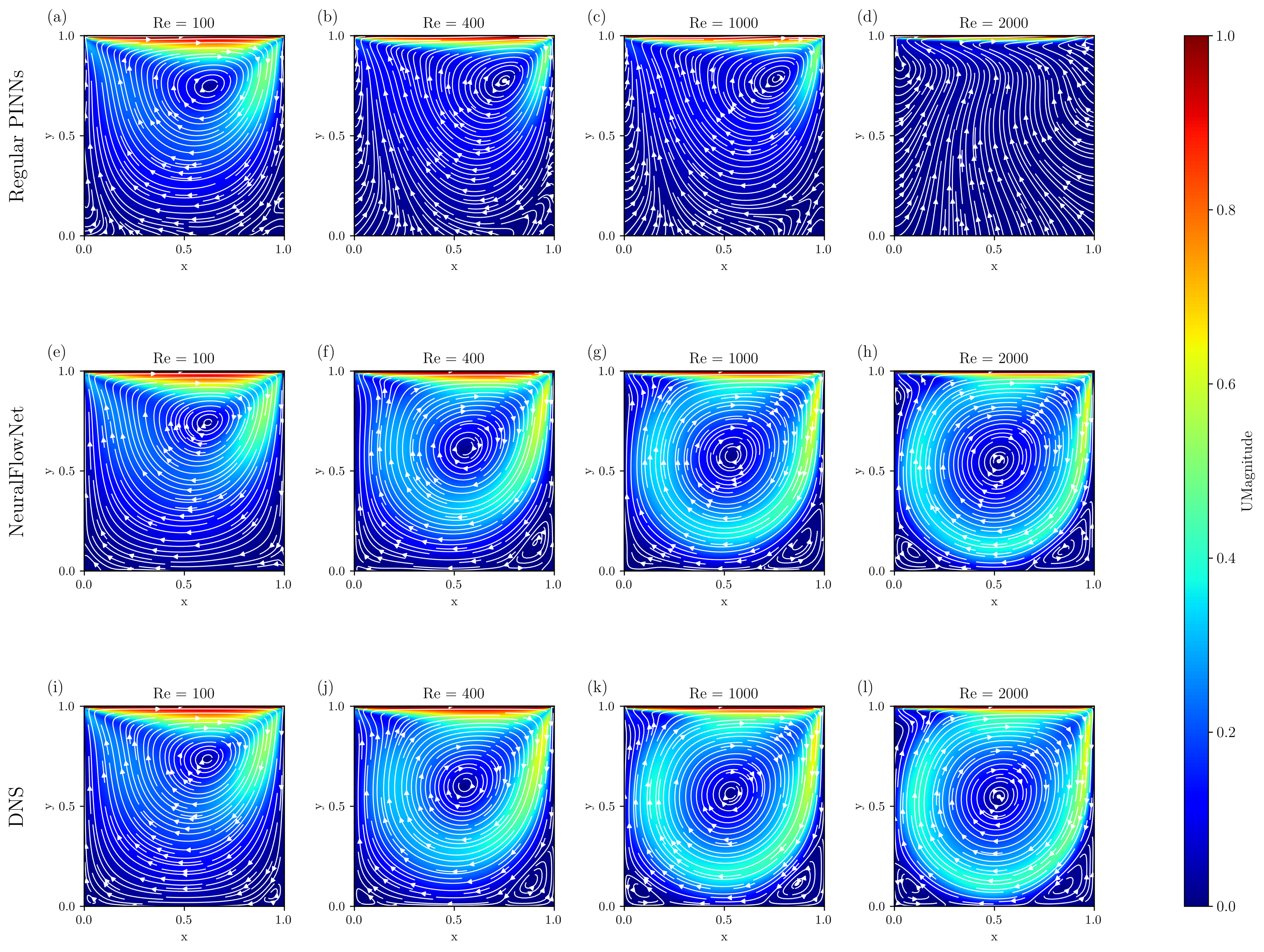}
    \caption{Comparison of cavity-flow velocity magnitude fields across Reynolds numbers Re=100, 400, 1000, and 2000 for Regular PINNs, \textit{NeuralFlowNet}, and DNS.}
    \label{fig2}
\end{figure*}

\textit{NeuralFlowNet} performance was assessed by comparing its predictions with the DNS reference solution at identical spatial locations. However, the best practices for solving many CFD problem consider the use of non uniform grid near the boundary, and following the contour. In our case, this was the approach used in the DNS channel, periodic hill, and pipe experiments. However, following exactly the implementation that openFoam uses in \textit{NeuralFlowNet}, especially during refining of the grid, it is not straightforward. For that reason, the training collocation points used by \textit{NeuralFlowNet} do not necessarily match exactly all the locations in the DNS mesh. However, all the results presented in this section were evaluated directly at the DNS coordinates, meaning that some of them are interpolated by the \textit{NeuralFlowNet}. We know that this evaluation could be slightly unfair with our implementation; however, this pointwise evaluation enables a consistent comparison with the reference flow fields, ensuring that the reported error metrics are comparable.

\subsection{Lid Driven Cavity}
As the first benchmark case, we selected the lid-driven cavity flow, a common CFD validation problem characterized by simple geometry and well-defined boundary conditions. This experiment retains the essential flow physics that any reliable solver must reproduce. In particular, this benchmark tests the ability of the model to capture the primary recirculation cell, the secondary corner eddies, the shear-driven cavity motion, and the associated pressure distribution. These features make the lid-driven cavity an appropriate first test for assessing the physical consistency and predictive capability of \textit{NeuralFlowNet} before extending the framework to more complex flow configurations. Following the classical configuration of \cite{ghia1982high}, both the OpenFOAM reference model and the \textit{NeuralFlowNet} computational domain were constructed on a \(129 \times 129\) grid. \textit{NeuralFlowNet} employed six hidden layers with 128 neurons per layer and used the SiLU activation function.

\begin{figure*}[!t]
    \centering
    \includegraphics[width=0.95\textwidth]{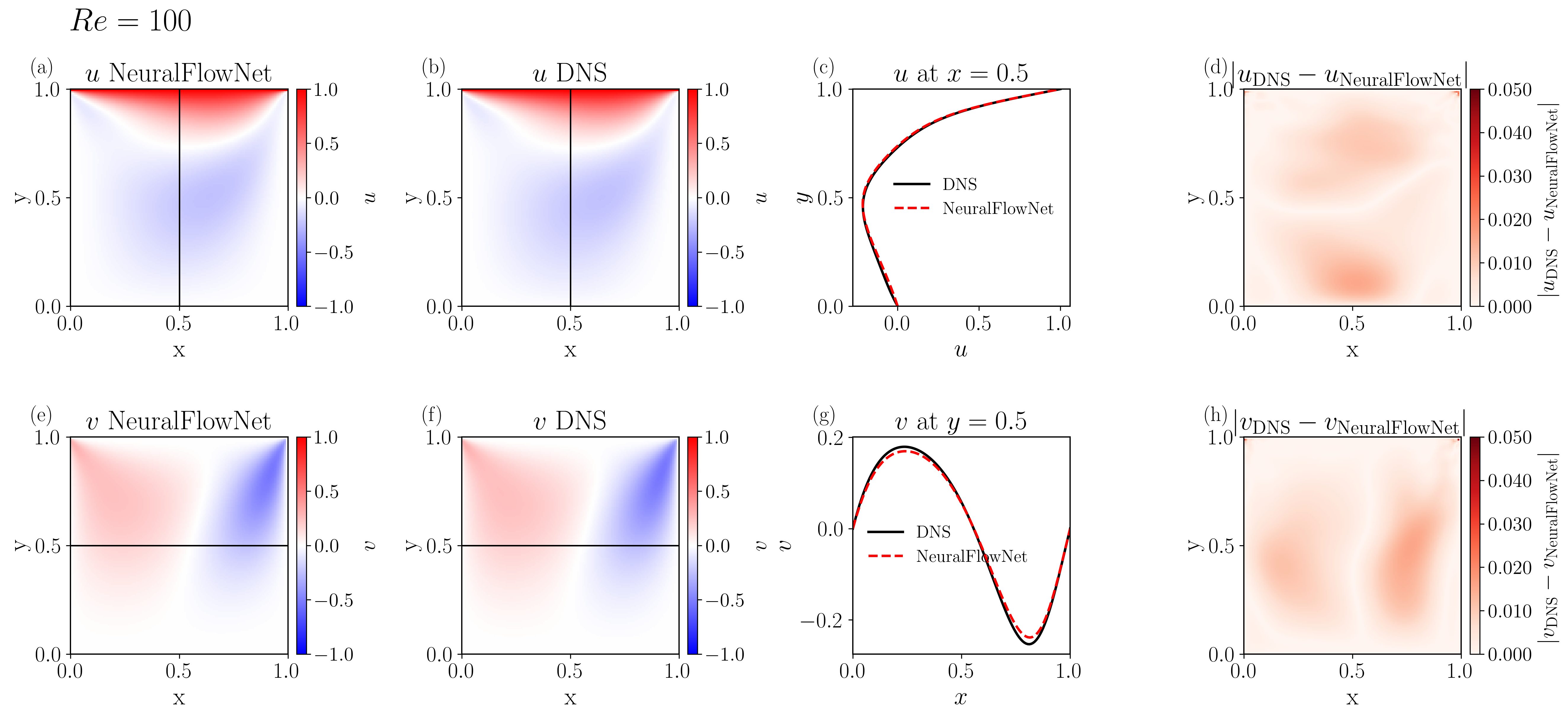}

    \vspace{0.3cm}

    \includegraphics[width=0.95\textwidth]{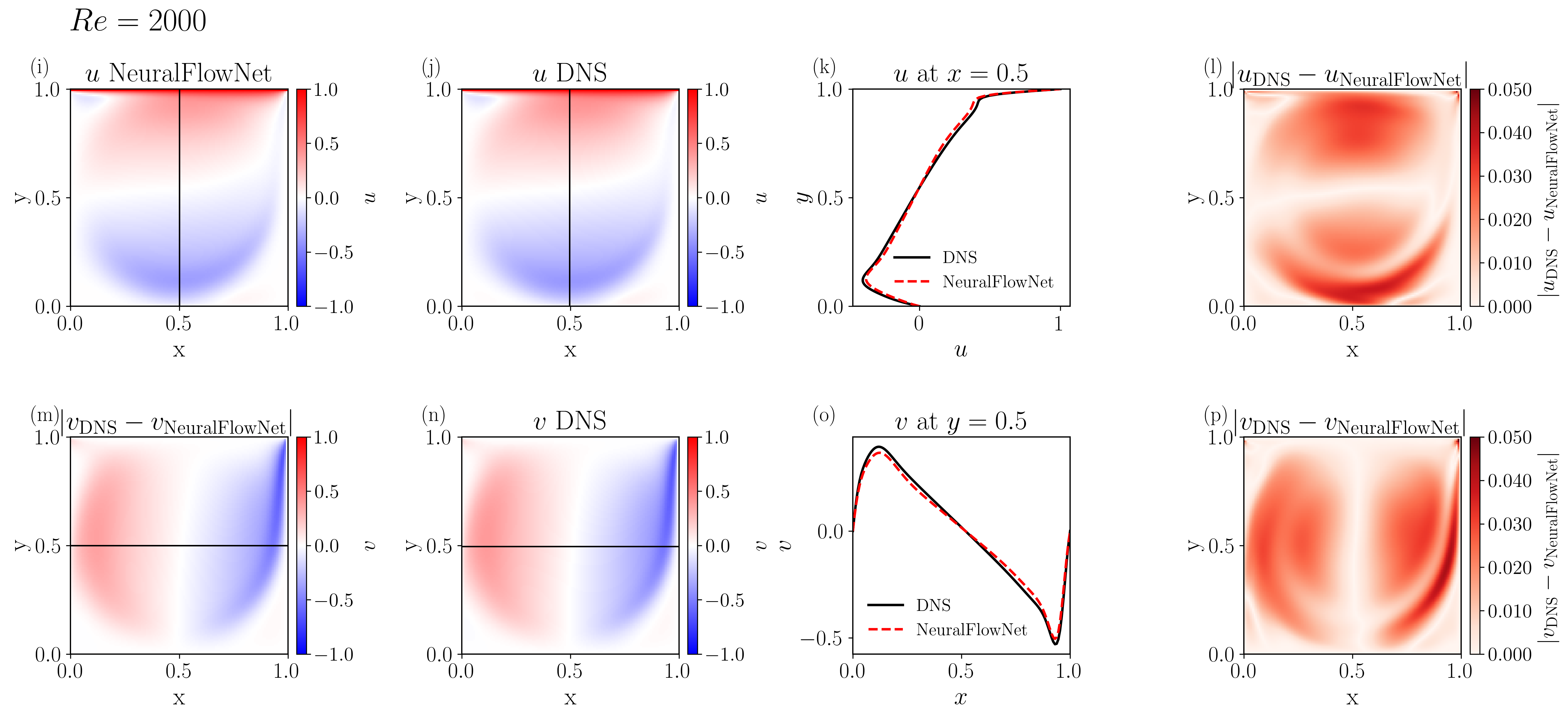}

    \caption{Comparison of lid-driven cavity flow results for Re = 100 and 2000. For each Reynolds number, the panels present the horizontal velocity component $u$ predicted by \textit{NeuralFlowNet}, the corresponding DNS solution, the centerline profile comparison, and the absolute difference, followed by the same set of comparisons for the vertical velocity component $v$.}
    \label{fig2}
\end{figure*}

We first compared Regular PINNs (without observation data and with AD), \textit{NeuralFlowNet}, and DNS reference solutions over a range of low to high Reynolds numbers (Figure~\ref{fig2}). The results show that the Regular PINNs failed to produce physically consistent solutions beyond \(Re = 100\), and even at \(Re = 100\), the predicted streamlines did not properly follow the expected flow structure. Regular PINNs tends to a solution space that does not allow tracking the evolution of high Reynolds numbers. In contrast, \textit{NeuralFlowNet} successfully reproduced the flow patterns across the full Reynolds-number range considered, showing close qualitative agreement with the DNS solutions. In particular, \textit{NeuralFlowNet} was able to capture both the primary vortex and the secondary corner eddies (Figure~\ref{fig2}h and l, bottom corner), indicating that the framework correctly resolved the dominant flow motion and recirculation behavior.

\begin{figure*}[!b]
    \centering
    \includegraphics[width=0.95\textwidth]{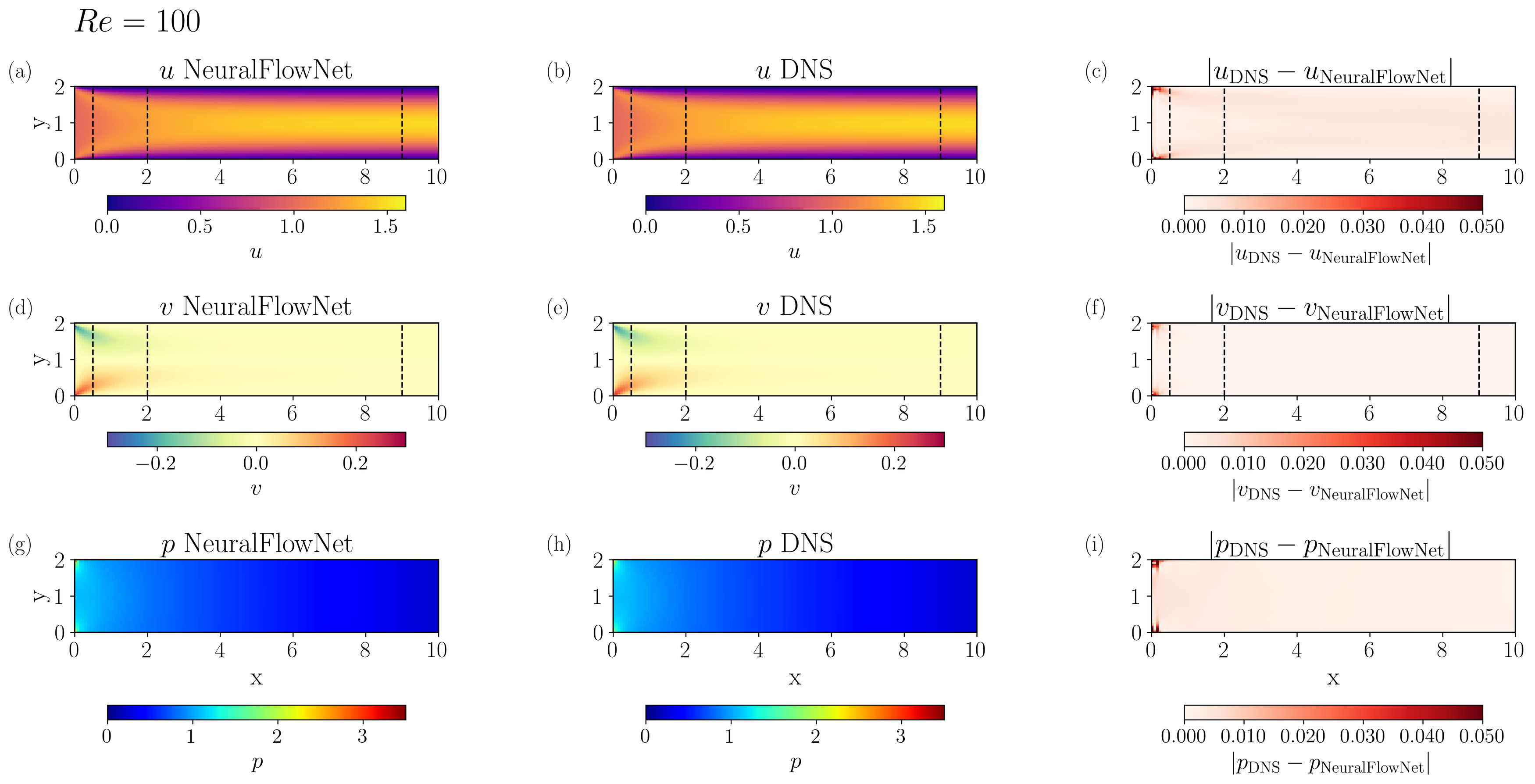}

    \vspace{0.3cm}

    \includegraphics[width=0.95\textwidth]{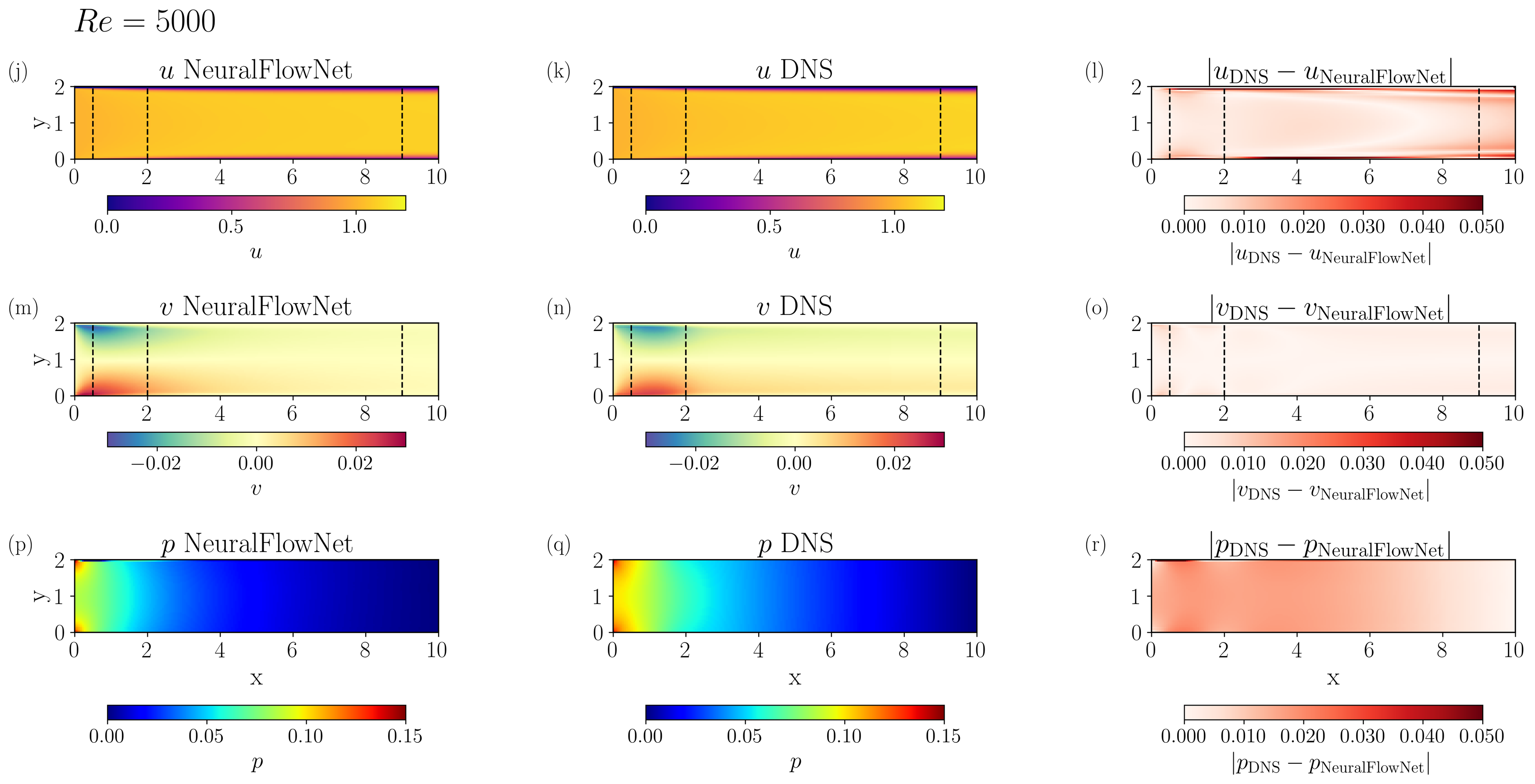}

    \caption{Comparison of channel flow results for Re = 100 and 5000. For each Reynolds number, the panels present the horizontal velocity component $u$ predicted by \textit{NeuralFlowNet}, the corresponding DNS solution, and the absolute difference, followed by the same set of comparisons for the vertical velocity component $v$ and $p$.}
    \label{fig3}
\end{figure*}

For further evaluation, we selected \(Re = 100\) and \(Re = 2000\) for detailed comparisons of the \(u\)- and \(v\)-velocity fields against the DNS reference solution, together with the corresponding absolute difference distributions. Table \ref{tab:benchmark_summary} reports the \(NRMSE\). In addition, two profile lines, \(x = 0.5\) and \(y = 0.5\), shown in Figure~\ref{fig2} (third column), were used to compare the predicted \(u\) and \(v\) velocity profiles with those from the DNS solution. At \(Re = 100\), \textit{NeuralFlowNet} reproduced both velocity components closely, with generally low absolute differences of approximately 0.02 (-), while the largest discrepancies were observed near the boundaries, likely due to the use of hard boundary constraints. The profile comparisons further confirmed that the predicted velocity distributions closely matched the DNS solution. At \(Re = 2000\), the model remained consistent in reproducing both \(u\) and \(v\), although the absolute difference increased slightly relative to the lower Reynolds-number case (0.05). Nevertheless, the velocity profiles continued to show strong agreement with the DNS reference. Overall, these results demonstrate that \textit{NeuralFlowNet} can reliably recover the dominant flow structures and velocity fields for this benchmark across a broad range of Reynolds numbers.
\subsection{Channel Flow}

\begin{figure*}[!h]
    \centering
    \includegraphics[width=0.95\textwidth]{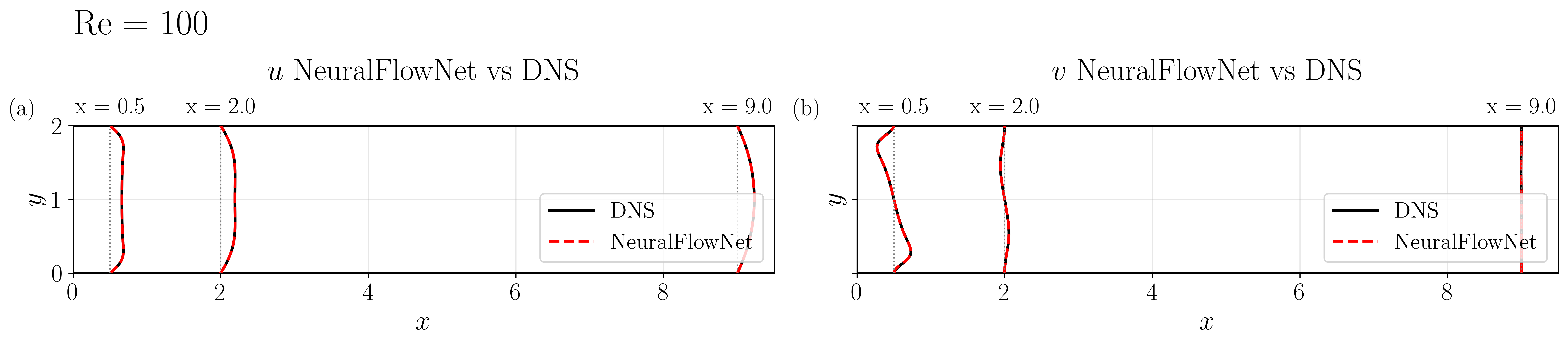}

    \vspace{0.3cm}

    \includegraphics[width=0.95\textwidth]{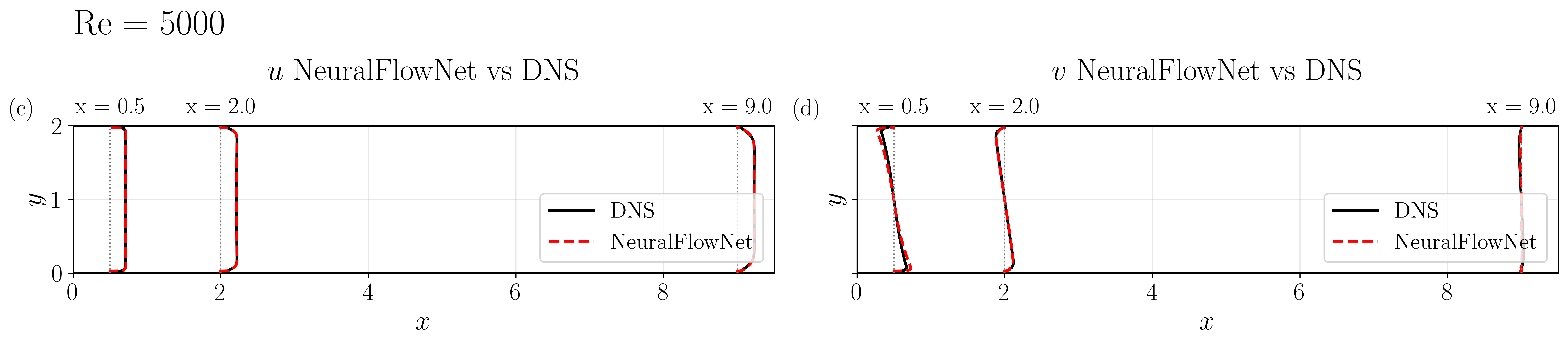}

    \caption{Comparison of streamwise ($u$) and transverse ($v$) velocity profiles at x=0.5, 2, and 9 for flow fields at Re=100 and Re=5000. (a) $u$ at Re=100, (b) $v$ at Re=100, (c) $u$ at Re=5000, and (d) $v$ at Re=5000.}
    \label{fig5}
\end{figure*}

As the second benchmark case, we considered a pressure-driven channel (Figure \ref{fig3}) flow formulated using a body-force approach. In contrast to the lid-driven cavity problem, which is driven by wall motion, channel flow is sustained by a constant streamwise forcing term and therefore provides a complementary test of the proposed framework under internal-flow conditions. Owing to its simple geometry, well-defined boundary conditions, and the availability of high-fidelity reference solutions, channel flow serves as an appropriate benchmark for assessing whether the model can accurately reproduce body-force-driven momentum transport, wall shear, viscous diffusion, and the associated pressure field.

For the channel-flow simulations, we retained the same \textit{NeuralFlowNet} neural network architecture used for the lid-driven cavity case, while introducing the streamwise body-force term and modifying the boundary conditions accordingly to represent the pressure-driven internal-flow configuration. The results show that \textit{NeuralFlowNet} successfully captured the flow behavior from low to high Reynolds numbers, with the predicted \(u\)- and \(v\)-velocity fields and pressure distribution showing close agreement with the CFD reference solutions. Both the absolute difference and the relative \(NRMSE\) remained low, indicating that the proposed framework can accurately reproduce the velocity structure, pressure response, and overall flow physics of pressure-driven channel flow, even better than the Lib Driven Cavity experiment.

Additionally, velocity profiles were evaluated at $x = 0.5$, $2$, and $9$ for both the $u$- and $v$-velocity components for the $Re = 100$ and $Re = 5000$ cases (Figure \ref{fig5}). For $Re = 100$, the predicted $u$- and $v$-velocity profiles at all three streamwise locations agree closely with the reference DNS results, demonstrating that the model accurately captures the flow dynamics under low-Reynolds-number conditions. For $Re = 5000$, the predicted $u$-velocity profiles remain in good agreement with the DNS data. The $v$-velocity profile shows a slight overestimation at $x = 0.5$; however, the profiles at $x = 2$ and $x = 9$ closely follow the DNS reference solution. Overall, these comparisons indicate that the model captures the main velocity structure across both Reynolds numbers, with only minor deviations in the near-inlet transverse velocity profile at the higher Reynolds number.

\subsection{Flow Over Periodic Hill}

\begin{figure*}[!h]
    \centering
    \includegraphics[width=1\textwidth]{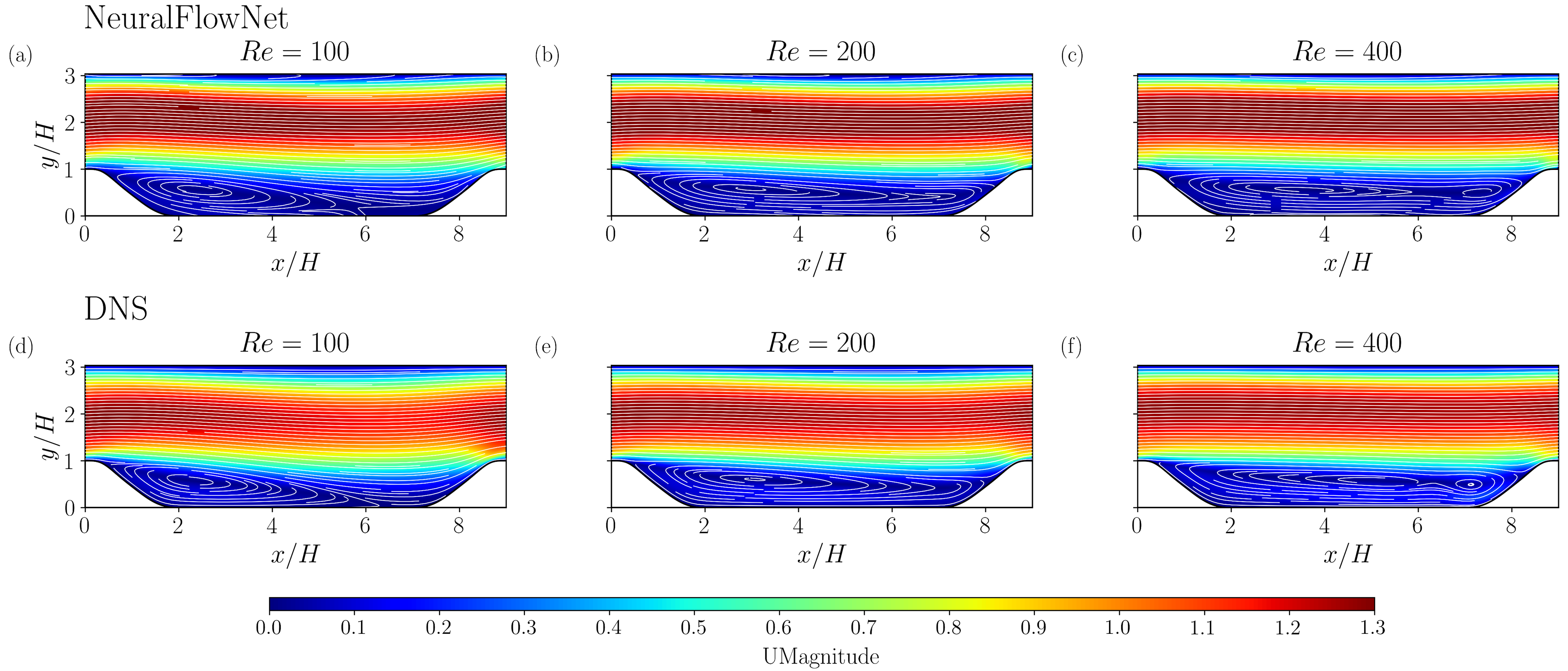}
    \caption{Comparison of flow over Periodic-Hill velocity magnitude fields across Reynolds numbers Re=100, 200 and 400, \textit{NeuralFlowNet}, and DNS.}
    \label{fig6}
\end{figure*}

As the third benchmark, we considered the 2D periodic-hill flow, which is widely used as a CFD validation case because it combines separated flow (See Figure \ref{fig6}), recirculation, and irregular wall geometry within a periodic domain \cite{breuer2009flow,rapp2011flow,greenshields_openfoam_2021}. This case was selected to assess whether \textit{NeuralFlowNet} can accurately represent flows over nontrivial curved boundaries, which is more demanding than the regular geometries of the lid-driven cavity and channel-flow benchmarks. In the present implementation, the hill geometry is represented explicitly through a body-fitted mesh, periodic boundary conditions are imposed in the streamwise direction, no-slip conditions are enforced at both the hill surface and the upper wall, and zero-gradient pressure conditions are applied at the lower and upper boundaries. Instead of prescribing a fixed mean-flow forcing term, the streamwise forcing was treated as a learnable scalar parameter and optimized together with the network so that the solution matched the target inlet bulk velocity (${U_b}$ = 1). These features make the periodic-hill case a useful test of both geometric flexibility and internal-flow forcing within the proposed framework.

\begin{figure*}[!b]
    \centering
    \includegraphics[width=1\textwidth]{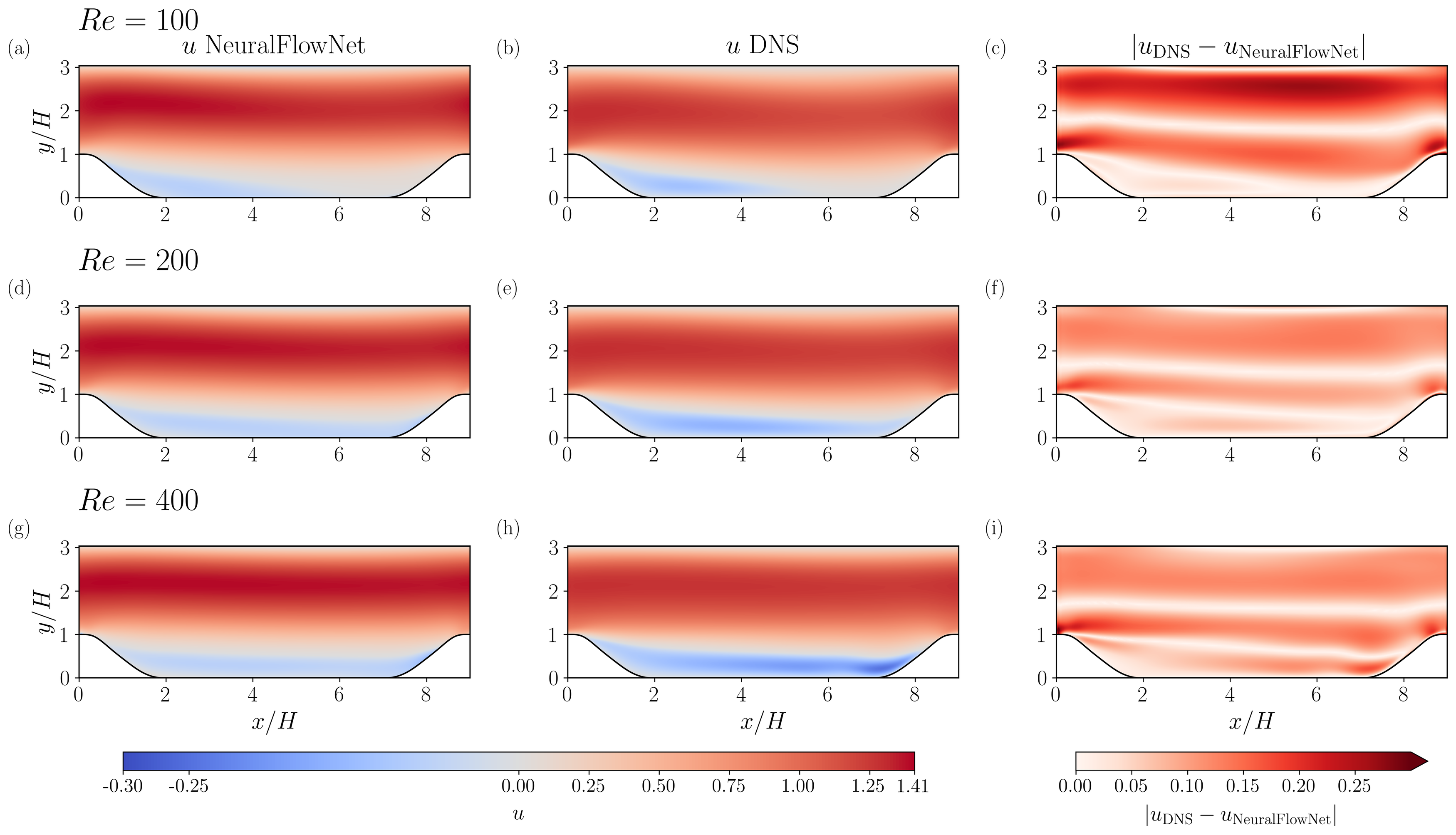}
    \caption{Comparison of flow over Periodic-Hill results for Re = 100, 200 and 400. For each Reynolds number, the panels present the horizontal velocity component $u$ predicted by \textit{NeuralFlowNet}, the corresponding DNS solution, and the absolute difference}
    \label{fig7}
\end{figure*}

\begin{figure*}[!h]
    \centering
    \includegraphics[width=1\textwidth]{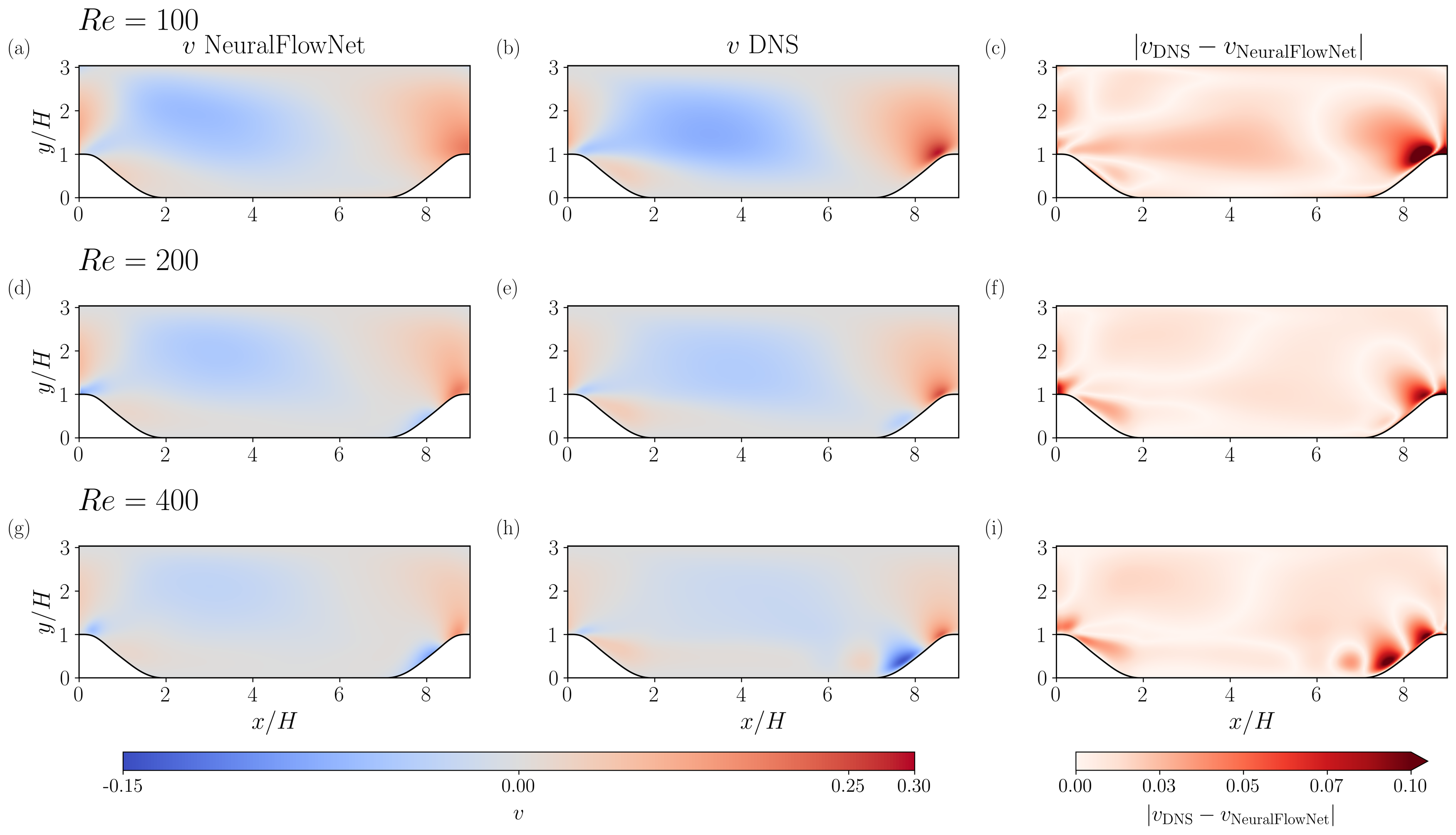}
    \caption{Comparison of flow over Periodic-Hill results for Re = 100, 200 and 400. For each Reynolds number, the panels present the vertical velocity component $v$ predicted by \textit{NeuralFlowNet}, the corresponding DNS solution, and the absolute difference}
    \label{fig8}
\end{figure*}

For the Reynolds-number cases $Re=100, 200,$ and $400$, \textit{NeuralFlowNet} reproduced the principal flow structures observed in the steady-state DNS solutions (See Figure \ref{fig6}). In particular, the velocity-magnitude contours and streamline patterns show comparable flow acceleration over the hill crest, flow separation, wake recirculation, and downstream reattachment. However, obtaining a converged steady-state solution using the DNS code was challenging, even with the prescribed convergence thresholds. This difficulty may be related to the recirculating flow and periodic boundary conditions, under which the outflow continuously re-enters the computational domain. In contrast, \textit{NeuralFlowNet} produces a steady-state solution by design. The network predicts spatial flow fields only, and the governing-equation residuals contain no time-derivative terms. Therefore, it directly seeks a stationary solution by minimizing the momentum, continuity, and boundary-condition residuals. Although \textit{NeuralFlowNet} can generate steady-state solutions at higher Reynolds numbers, the present comparison was limited to $Re \leq 400$, for which sufficiently converged steady-state DNS reference solutions were available. Despite differences in the numerical formulations and convergence criteria, the close agreement between the velocity-magnitude and streamline structures demonstrates that \textit{NeuralFlowNet} effectively captures the dominant flow features of the periodic-hill cases.

We compared the $u$- and $v$-velocity components predicted by \textit{NeuralFlowNet} with the steady-state DNS solutions (Figure \ref{fig7}). For the $u$-velocity component, the model reproduced the overall spatial distribution at all Reynolds numbers, including the high-velocity region in the upper domain and the negative velocities within the recirculation region. The domain-wide NRMSE values were 7.02\%, 5.05\%, and 5.27\% at $Re=100$, $200$, and $400$, respectively, indicating good overall agreement with DNS. The absolute difference show that differences exceeding 0.25 were localized. At $Re=100$, these differences occurred mainly in the upper domain, where positive velocities were slightly overpredicted. At $Re=200$, positive velocities were slightly overpredicted, while negative velocities were underestimated. Similar differences occurred at $Re=400$, primarily near the downstream hill, where DNS predicted stronger reverse flow. Despite these localized discrepancies, \textit{NeuralFlowNet} captured the dominant streamwise-flow behavior throughout most of the domain.

For the $v$-velocity component, \textit{NeuralFlowNet} reproduced the overall distribution of upwelling and downwelling flow at all Reynolds numbers (Figure \ref{fig8}). The domain-wide NRMSE values were 6.50\%, 6.17\%, and 4.97\% at $Re=100$, $200$, and $400$, respectively, indicating good overall agreement with DNS. The absolute differences remained below 0.1 across the domain and were predominantly below 0.05, with the largest values localized near the hill boundaries. At $Re=100$ and $200$, the model slightly underestimated the positive velocities near the downstream hill. At $Re=400$, the largest differences occurred along the downstream slope, where DNS predicted stronger vertical motion. Nevertheless, \textit{NeuralFlowNet} accurately reproduced the $v$-velocity field throughout most of the domain.

Overall, \textit{NeuralFlowNet} reproduced the dominant $u-$ and $v-$velocity distributions observed in the steady-state DNS solutions for all three Reynolds numbers. The model captured the main high-velocity, reverse-flow, and vertical-motion regions. The remaining differences were primarily localized within the separated-flow region and near the downstream hill, where strong velocity gradients and complex flow interactions occur. These discrepancies may partly result from differences in the numerical formulations, convergence behavior, and stopping criteria used by the DNS and \textit{NeuralFlowNet} approaches.

\begin{figure*}[!h]
    \centering
    \includegraphics[width=0.85\textwidth]{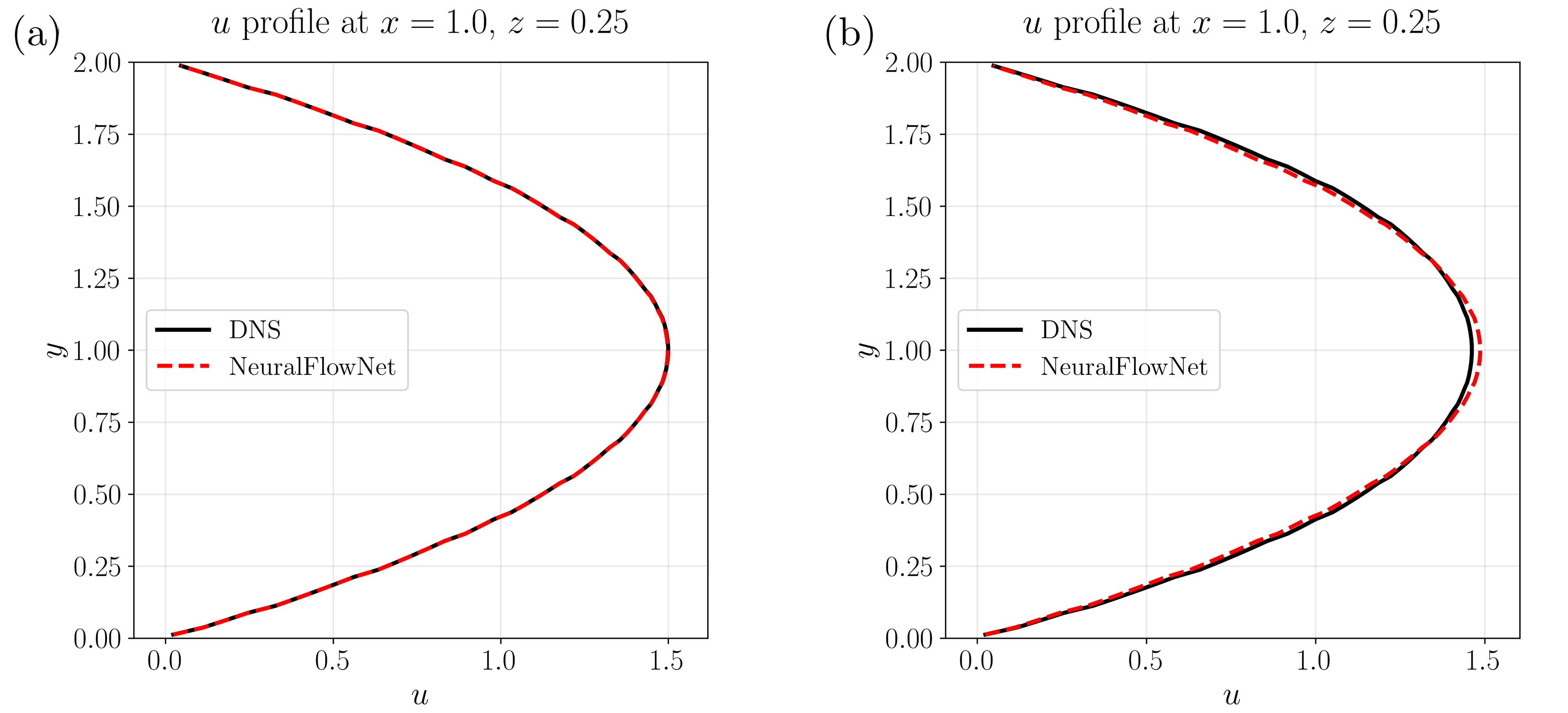}

    \caption{\textit{NeuralFlowNet} comparison the corresponding mid-channel $u$ profiles with the DNS reference data (a) $Re$=100 (b) $Re$= 3000}
    \label{fig9}
\end{figure*}

\subsection{3D Internal Flow Driven by a Constant Body Force}

As the fourth benchmark, we considered a 3D internal-flow case driven by a constant streamwise body force (Figure \ref{fig9}). This case was selected as a proof-of-concept step to extend \textit{NeuralFlowNet} from 2D configurations to a fully 3D setting before moving to more geometrically and dynamically complex 3D flows. The objective was to verify that the proposed framework can recover a 3D wall-bounded solution, preserve the dominant streamwise momentum transport, and maintain stable learning in a higher-dimensional domain. In the present implementation, the case was formulated as a laminar Poiseuille-type internal flow in a periodic domain. Unlike a conventional inlet--outlet pressure-driven configuration, the flow here is maintained by a constant streamwise body force, which serves as the periodic-domain analogue of a pressure gradient. This avoids the need to impose explicit inflow and outflow boundary conditions while still producing the expected internal-flow solution.

The predicted 3D fields showed close agreement with the reference solution for both Reynolds numbers considered. For $Re=100$ and $Re=3000$, it is difficult to distinguish the differences between the DNS and \textit{NeuralFlowNet} results (see Appendix). For $Re=100$, the absolute difference in the velocity magnitude remained on the order of $10^{-3}$, indicating very strong agreement throughout the domain. For $Re=3000$, the absolute difference in the velocity magnitude ranged from 0 to approximately 0.032, showing a moderate increase relative to the lower-Reynolds-number case while still preserving the overall flow structure. Therefore, we examined the cross-section at $z = 0.25$, taken normal to the $z$-direction, for both $Re=100$ and $Re=3000$ to further evaluate the ability of \textit{NeuralFlowNet} to recover the 3D internal-flow solution (see Figure \ref{fig10}). For this case, the analytical maximum velocity is known to be 1.5. The results show that \textit{NeuralFlowNet} reproduces this value correctly, whereas the DNS reference slightly underestimates it, indicating that the DNS solution may not have fully converged under the adopted stopping criterion. This further supports that \textit{NeuralFlowNet} is able to recover the expected flow dynamics accurately.

\subsection{3D Backward-Facing-Step Flow}

\begin{figure*}[!h]
    \centering
    \includegraphics[width=0.95\textwidth]{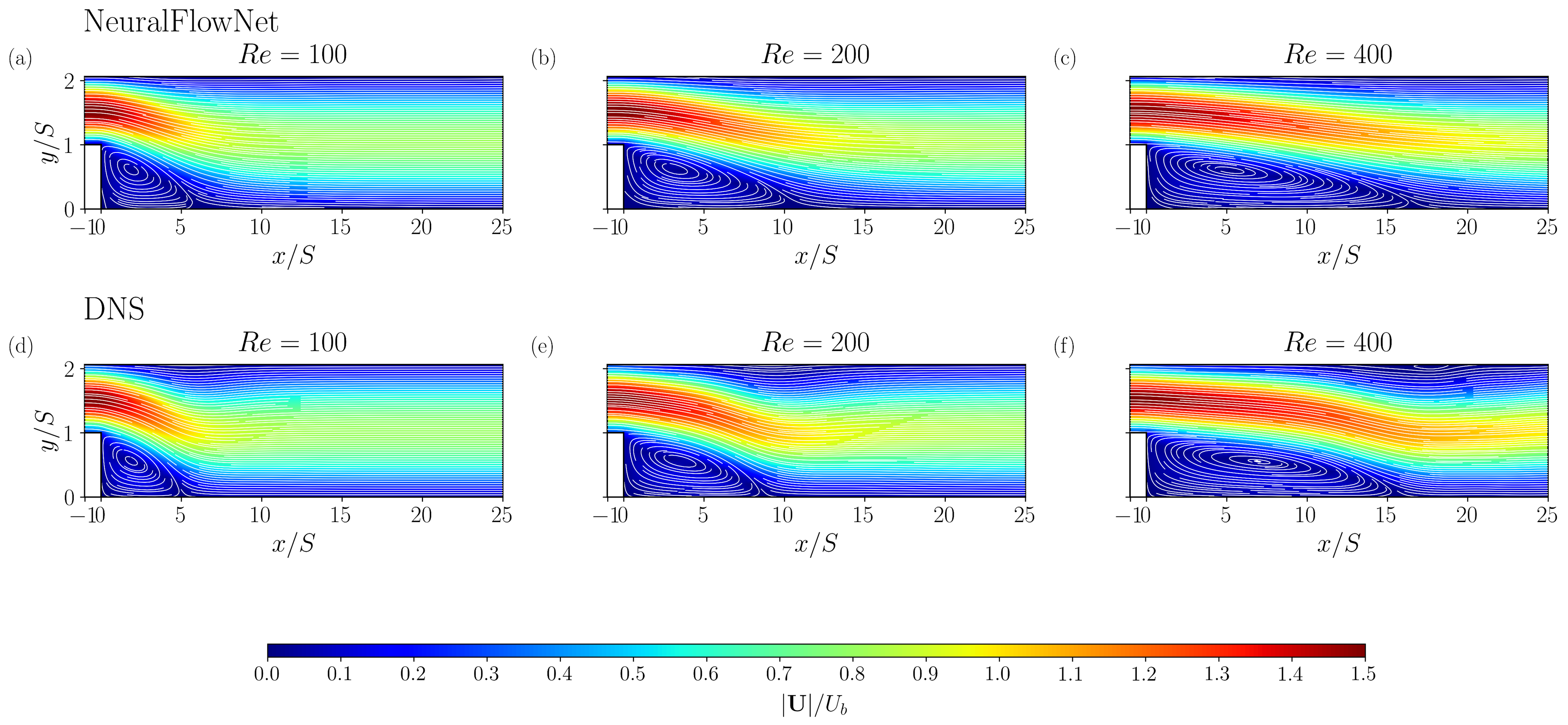}

    \caption{Comparison of the BFS velocity-magnitude at the midplane for Reynolds numbers Re=100, 200, and 400., \textit{NeuralFlowNet}, and DNS.}
    \label{fig10}
\end{figure*}

\begin{figure*}[!b]
    \centering
    \includegraphics[width=0.95\textwidth]{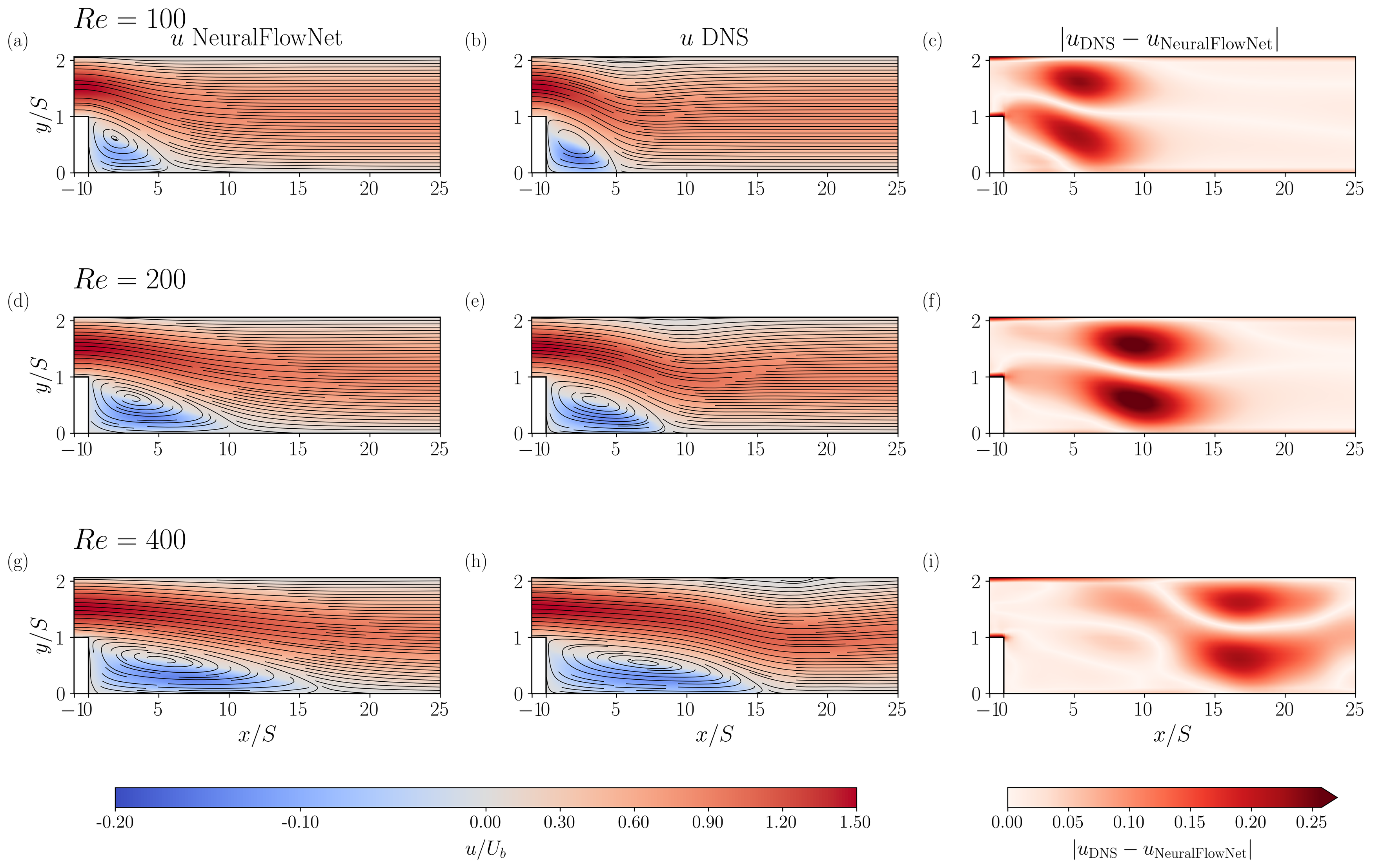}

    \caption{Comparison of the BFS $u$-velocity component at the midplane for Reynolds numbers Re=100, 200, and 400., \textit{NeuralFlowNet}, and DNS.}
    \label{fig11}
\end{figure*}

\begin{figure*}[!h]
    \centering
    \includegraphics[width=0.95\textwidth]{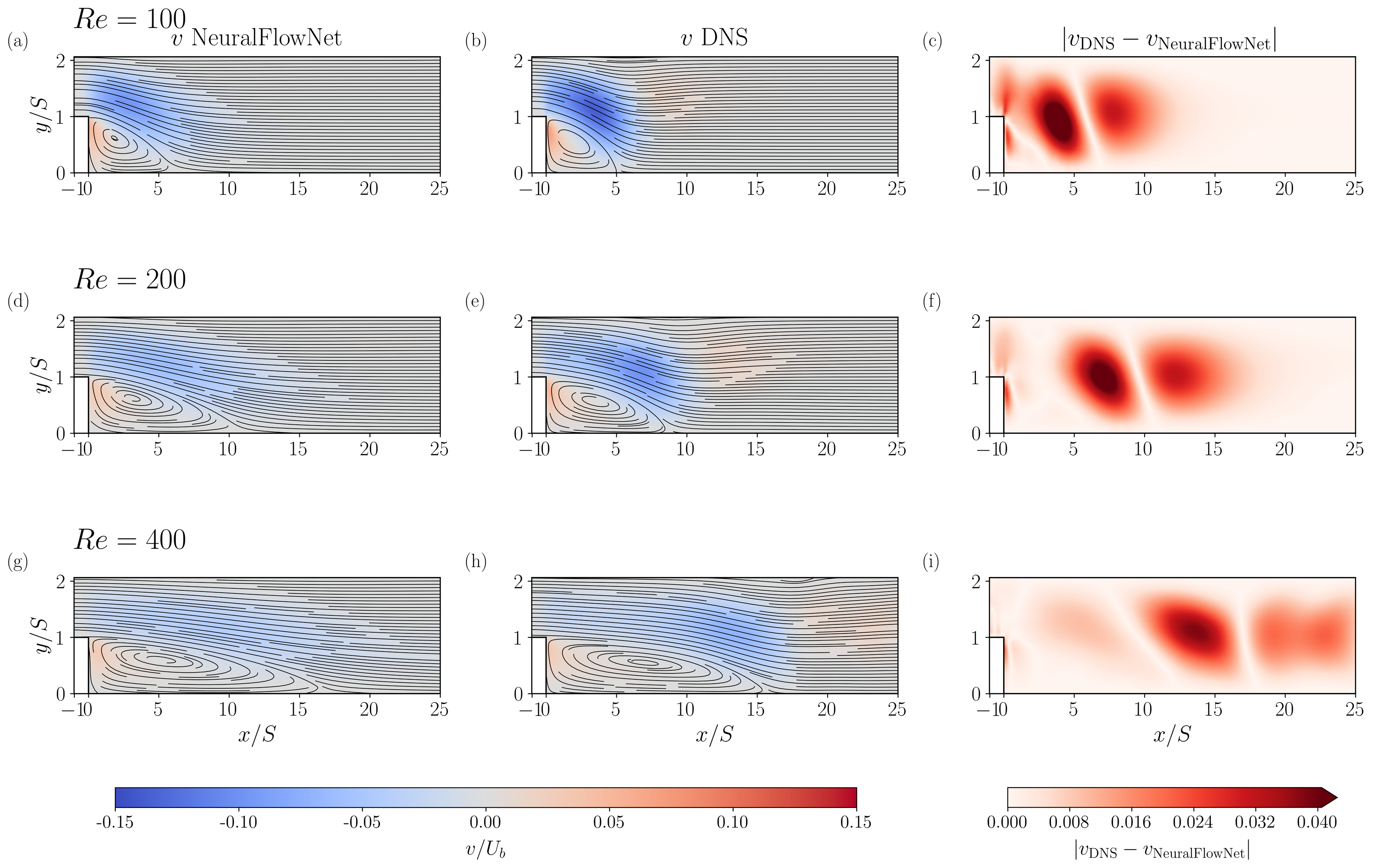}

    \caption{Comparison of the BFS $v$-velocity component at the midplane for Reynolds numbers Re=100, 200, and 400., \textit{NeuralFlowNet}, and DNS.}
    \label{fig12}
\end{figure*}

As a further three-dimensional benchmark, we considered flow over a backward-facing step (BFS) (Figure \ref{fig10}). This case is substantially more demanding than the preceding 3D internal-flow proof of concept because it involves an abrupt geometric expansion, flow separation, the formation of a recirculation region, and downstream reattachment. The computational domain was based on the geometry employed in the numerical study of \cite{williams1997numerical} and was nondimensionalized using the step height ($S$). A fully developed laminar velocity profile for a rectangular duct, normalized to a bulk velocity of (\(U_b=1\)), was prescribed at the inlet. No-slip conditions were imposed on the channel walls, the step surfaces, and the outer spanwise wall. At the outlet, a zero-normal-gradient condition was applied to the velocity, while the pressure was fixed to zero. A symmetry condition was imposed at the spanwise center plane. The analysis considered Reynolds numbers in the range \(100\le Re\le400\). This range was selected to examine the evolution of the separated flow while maintaining a computationally feasible 3D low Reynolds number benchmark. Because the reattachment location is expected to move downstream as \(Re\) increases, the downstream domain length of \(30.61\) was retained to reduce the influence of the outlet boundary on the recirculation region. The results were additionally checked to confirm that reattachment occurred sufficiently upstream of the outlet.

The \textit{NeuralFlowNet} solutions demonstrate that the model successfully recovers the principal flow structures of the BFS configuration. In particular, the model captures flow separation at the step and the subsequent recirculation region across all Reynolds numbers. The predicted reattachment locations are approximately $x/S=5$, $10$, and $15$ for $Re=100$, $200$, and $400$, respectively. These predictions agree closely with the corresponding DNS results, although they are slightly overpredicted (Figure~\ref{fig10}). The progressive elongation of the recirculation region with increasing Reynolds number is also reproduced accurately. Although slight differences are observed in the normalized velocity-magnitude distributions (\(|U|\)/\(U_b\)), the overall spatial structure and magnitude of the flow remain consistent with the DNS solutions.

Velocity component-wise comparison further shows that the spatial distribution of the normalized streamwise velocity, (\(u\)/\(U_b\)), is captured well by \textit{NeuralFlowNet} (Figure \ref{fig11}). The model reproduces both the negative streamwise velocities within the recirculation region and the high-velocity region above the separated shear layer. However, \(u\)/\(U_b\) is slightly overestimated by \textit{NeuralFlowNet}. Thus, the largest absolute differences occur primarily along the separated shear layer and within the downstream high-velocity region, whereas relatively small differences are observed inside much of the recirculation zone. This discrepancy becomes more spatially extensive and shifts downstream as the Reynolds number increases.

For the normalized wall-normal velocity, \(v\)/\(U_b\)), both \textit{NeuralFlowNet} and DNS show stronger downwelling motion at $Re=100$, followed by a progressive reduction as the Reynolds number increases (See Figure \ref{fig12}). \textit{NeuralFlowNet} slightly overpredicts the magnitude of this downward velocity, particularly near the separated shear layer immediately downstream of the step. Accordingly, the largest absolute difference is observed at $Re=100$, while its magnitude and spatial extent decrease at $Re=200$ and $400$. Overall, these comparisons indicate that \textit{NeuralFlowNet} accurately captures the Reynolds-number-dependent evolution of the separated flow, despite localized deviations in the streamwise and wall-normal velocity magnitudes.

\section{Discussion}

The present study shows that \textit{NeuralFlowNet} can recover steady incompressible flow solutions in both 2D- and 3D benchmark problems without relying on training data. This is an important result for both the CFD and ML communities. From the CFD perspective, it demonstrates that physically meaningful flow structures, including recirculation, reattachment, and wall-bounded transport, can be recovered by a neural framework when the governing operators are enforced in a numerically consistent way. This way, \textit{NeuralFlowNet} can set the basis for the development of an alternative method for fully solving the NS equations. From the ML perspective, it shows that data-free scientific learning can be made substantially more robust when the network is guided not only by the governing equations, but also by the numerical method through which \textit{those equations are traditionally solved}. This creates a new perspective on what a really Physics-Informed Neural Network is. Now, decades of knowledge about how to deal with the intrinsic complexities of the NS equations is getting value, which help to fill in the gaps between the CFD and ML communities. 

It is important to emphasize that the DNS results are used as high-fidelity reference solutions rather than exact ground truth. DNS provides an appropriate basis for comparison because it solves the NS equations directly, without turbulence modeling, using established spatial discretization and pressure--velocity coupling procedures. It therefore provides an independent and widely accepted benchmark for evaluating whether \textit{NeuralFlowNet} recovers the expected flow structures and velocity distributions. The DNS cases were computed in OpenFOAM using \texttt{simpleFoam}, with the steady fields obtained iteratively. Their accuracy depends on the grid resolution, spatial discretization, residual convergence, and stopping criteria. For some cases, the velocity and pressure residuals approached but did not fully satisfy the prescribed convergence thresholds, despite a large number of iterations. These results should therefore be regarded as approximately converged steady DNS solutions.\textit{NeuralFlowNet}, in contrast, directly approximates the solution of the steady NS equations, for which the temporal terms are omitted. Its solution is affected by spatial discretization, neural representation, and optimization errors, but not by temporal discretization errors. Consequently, discrepancies between the two methods cannot be interpreted solely as differences in \textit{NeuralFlowNet}. Rather, the comparison assesses whether two independent numerical formulations recover consistent solutions to the same governing equations and boundary conditions. The DNS results remain the reference because they were obtained using a mature and extensively validated CFD methodology, while \textit{NeuralFlowNet} is the numerical framework being evaluated.

The significance of this work also extends beyond the benchmark cases considered here. At present, much of the machine-learning activity in fluid mechanics is focused on surrogate modeling (copying other model behavior), reduced-order prediction (simplified version of other model), or data-driven reconstruction (mimicking the behavior learned from data). In contrast, \textit{NeuralFlowNet} is formulated as a solver that learns directly from the governing equations and boundary conditions. This gives it a different role from conventional supervised ML models. Rather than learning from examples alone, it learns from the structure of the physical problem itself. For the CFD community, this suggests a pathway toward AI-based solvers that remain interpretable and physically constrained. For the ML community, it shows that incorporating subject-specific numerical method can significantly broaden the applicability of data-free learning.

From our development stage process, is have been clear that the main limitation of classical PINNs in fluid dynamics is not simply the lack of data, but the way in which the physics is imposed. In standard PINNs, the governing equations are evaluated pointwise using AD. Although AD provides exact derivatives of the neural-network representation, it does not explicitly incorporate the neighboring-point interactions that are fundamental to continuum transport. In fluid flow, the state at a given location is strongly influenced by the surrounding field through nonlinear advection, pressure coupling, viscous diffusion, and incompressibility. As the Reynolds number increases, these couplings become increasingly sensitive because shear layers, near-wall gradients, and separated-flow structures require an accurate local transport balance. Moreover, PINNs approximate a continuous function representing the solution throughout the computational domain. Increasing the number of neurons increases the network’s representational capacity. However, our tests suggest that network size alone does not explain the observed difficulties. The issue may instead be related to the effective degrees of freedom in the learned first- and second-order derivatives. The first derivatives represent advection and pressure gradients. The second derivatives represent viscous diffusion. Currently, there is no established method for determining the appropriate complexity of these derivative fields. A predicted solution may appear accurate and smooth, while its pointwise derivatives still fluctuate rapidly. This issue may be particularly important for second-order derivatives. Piecewise, discretization-inspired derivative operators may reduce this sensitivity. These operators couple neighboring solution points when evaluating the governing equations. Although this explanation remains speculative, it provides a direction for future research. Further work is needed to determine the appropriate complexity of learned derivative fields and evaluate their effects on conventional PINNs.

The present framework addresses this barrier by replacing purely pointwise residual evaluation with CFD-inspired discrete operator treatment. In \textit{NeuralFlowNet}, the governing terms are evaluated using finite difference in 2D and finite volume in 3D, so that neighboring-cell influence enters the residual directly. Moreover, it is important to consider the implications of such piecewise approximations and the specific behaviors that are involved in the NS resolution. The advection term is highly sensitive to stability and artificial numerical diffusion, the pressure-gradient term must remain symmetric and unbiased, the viscous term must preserve second-order smoothing behavior, and the divergence-free constraint must be enforced consistently across the domain. Treating all of these terms through a single generic neural approximation is therefore insufficient. The improved performance observed here suggests that a successful data-free PINN solver for fluid dynamics must be not only physics-informed, but also numerically method informed.

A second important observation concerns the role of the neural representation itself. In PDE learning, both first- and second-order derivatives enter directly into the residuals. That requires that the selected activation function must stably support both. Our experiments showed that commonly used activations (e.g., tanh) do not perform equally well in this setting. Among the functions examined, SiLU and $x\tanh(x)$ provided the most reliable behavior, suggesting that derivative quality is a central consideration for PDE-oriented neural solvers. This point is especially relevant to the ML community because it indicates that successful scientific machine learning cannot rely only on generic deep-learning design functions but must also account for the ease (or improvement) provided by other activation functions and for the specific structures needed for different PDE equations.

The third distinction concerns the spatial representation of the solution. In DNS, the governing equations are discretized and solved on a prescribed computational mesh. The resulting flow variables are defined at the corresponding cell centers, nodes, or faces. A solution on a different mesh generally requires remeshing and repeating the simulation. In \textit{NeuralFlowNet}, the trained neural network defines the flow variables as continuous nonlinear functions of the spatial coordinates. The solution can therefore be evaluated at arbitrary locations after training. It can also be sampled on grids that differ from the training grid without resolving the governing equations. In the present study, the \textit{NeuralFlowNet} training grids and DNS meshes were not identical. For the quantitative comparison, however, the trained network was evaluated at the DNS sampling locations. This provided a consistent basis for pointwise evaluation. The resulting capability should be described as grid-independent solution sampling rather than grid-independent accuracy. Sampling the network on a finer grid produces a more densely resolved representation of the learned solution. It does not necessarily add physical scales that were unresolved during training. The effective resolution remains controlled by the training grid, network capacity, discrete operators, and optimization accuracy.

Despite all the progress present in this research, several limitations remain. At the moment, the current framework is restricted to steady-state flow problems, and transient flows remain an open challenge. However, we consider that the extension of \textit{NeuralFlowNet} to transient flow problems is completely feasible, where temporal transport and evolving coherent structures must be represented in addition to spatial coupling. Second, the three-dimensional cases are still computationally expensive, particularly when combined with multistage training, adaptive control, and grid refinement; therefore, more experimentation must be done to create a highly efficient AI-solver that uses all the benefits of the new AI technology. Third, some controller thresholds and adaptive coefficients were selected empirically through numerical calibration. Although these settings produced robust convergence for the present benchmark cases, more experimentation with different scenarios could refine their values, which would improve the generality of the framework.

Another future line of exploration could address the need for designing activation functions, optimizers, and controller strategies specifically tailored for PDE learning. More broadly, the present work suggests that the lessons learned here are not limited to fluid dynamics. Any scientific problem governed by partial differential equations may benefit from a framework in which machine learning is constrained not only by the physics but also by the numerical character of the governing operators, in which the previous knowledge coming from scientists can play a fundamental role.

\section{Conclusions}

In this work, we developed \textit{NeuralFlowNet}, a data-free neural framework for solving the steady incompressible NS equations. The main motivation was to address the persistent limitations of classical PINNs in fluid-flow problems. Particularly, their difficulty in recovering accurate solutions at moderate and high Reynolds numbers without supervised data. To overcome these issues, we proposed a CFD-inspired framework that combined hard boundary conditions, Finite Difference and Finite Volume style implementation, adaptive treatment of advection and diffusion, projection-based incompressibility correction, and staged multilevel training. As a result, \textit{NeuralFlowNet} was formulated as a solver that is not only physics-guided, but also numerically informed.

The benchmark results showed that \textit{NeuralFlowNet} can recover meaningful steady incompressible-flow solutions in both 2D and 3D. The framework reproduced the main flow fields for the lid-driven cavity and channel-flow cases, captured irregular geometry and separated-flow behavior in the periodic-hill benchmark, and successfully extended to 3D internal-flow and BFS configurations. In particular, the method was able to recover recirculation and reattachment behavior in the more demanding separated-flow cases, demonstrating improved robustness compared with classical data-free PINNs.

One of the main limitations of classical PINNs in fluid dynamics arises not only from the absence of data, but also from the pointwise manner in which the governing equations are enforced. By introducing neighboring-point influence directly into the residual construction, \textit{NeuralFlowNet} shows that data-free scientific ML can be made significantly more effective for PDE-governed problems. 

Overall, the main contribution of this work is not only in the creation of a new data-free neural solver for fluid dynamics, but also a broader methodological insight relevant to both CFD and ML. In this sense, the present study argues that for scientific machine learning to become genuinely reliable in PDE problems, it must be built at the intersection of physics, computational methods, and machine learning rather than from any one of these components alone.

Although the present results are promising, important limitations remain. The framework is currently restricted to steady-state flows, the 3D cases are still computationally expensive, and some controller settings could still be finetuned. However, we believe that they are completely reachable in future work, which could convert \textit{NeuralFlowNet} into a completely alternative method for solving Navier-Stokes equations.

\section*{Acknowledgments}
This research was supported in part by the National Science Foundation (NSF) through the CAREER Award No.~2239550 and the NSF Implementation Grant ``Community-driven Inclusive Excellence and Leadership Opportunities in the Geosciences (CIELO-G)'' Award No.~2228180, and by the Army Research Office (ARO) under Award No.~226351995A.

The views and conclusions contained in this document are those of the authors and should not be interpreted as representing the official policies, either expressed or implied, of the ARO, NSF for the US Government. The US Government is authorize and distribute reprints for Government purposes notwithstanding any copyright notation herein.

\section*{AUTHOR DECLARATIONS}
Conflict of Interest\\
The authors have no conflicts to disclose.

\section*{Author Contributions}
Author Contributions: Conceptualization, J.T.S. and L.D.F; methodology, J.T.S, and L.D.F; software, J.T.S.; formal analysis, J.T.S. and L.D.F; resources, J.T.S., and L.V.A.; data curation, J.T.S.; writing--original draft preparation, J.T.S.; writing--review and editing, J.T.S, L.D.F and L.V.A; visualization, J.T.S.; supervision, L.D.F. and L.V.A; project administration, J.T.S. and L.V.A. All authors have read and agreed to the published version of the manuscript.

\section*{DATA AVAILABILITY}
The data that support the findings of this study are available
from the corresponding author upon reasonable request

\section*{REFERENCES}

\printbibliography[heading=none]

\newpage

\clearpage
\section*{Appendix}

\begingroup
\footnotesize

\setlength{\abovedisplayskip}{5pt}
\setlength{\belowdisplayskip}{5pt}
\setlength{\abovedisplayshortskip}{3pt}
\setlength{\belowdisplayshortskip}{3pt}

\begin{center}
\begin{minipage}{0.97\textwidth}

\hrule
\vspace{0.15cm}

\textbf{Algorithm 1:} Training procedure of
\textit{NeuralFlowNet}

\vspace{0.15cm}

\textbf{Input:} $\Omega$, BCs, $\{Re_m\}$, $\{N_s\}$,
$\{E_s\}$, $\{\eta_s\}$
\hfill
\textbf{Output:} $\theta^{*}$

\begin{enumerate}[leftmargin=*, itemsep=2pt, topsep=4pt]

     \item Initialize the network parameters $\theta$ and define the MLP predictor
    \[
    (\mathbf{u},p)=\mathcal{N}_{\theta}(\mathbf{x}),
    \]
    where $\mathbf{x}\in\mathbb{R}^{d}$ and $d=2$ or $3$.

    \item For each Reynolds number $Re_m$ and training stage $N_s$, construct the computational grid, transfer converged weights from the previous stage or Reynolds number if available, predict the primitive variables over $\Omega$, impose hard Dirichlet and Neumann boundary conditions, and normalize the pressure by
    \[
    p \leftarrow p-\frac{1}{N_{\Omega}}\sum_{\Omega} p.
    \]

    \item Compute the normalized epoch coordinate
    \[
    t=\frac{ep}{\max(E-1,1)}
    \qquad \text{(Eq.~\ref{eq8})}.
    \]

    \item Evaluate the discrete operators. Compute the blended advection term
    \[
    \mathcal{A}=(1-\beta)\mathcal{A}_{\mathrm{upwind}}+\beta\mathcal{A}_{\mathrm{central}}
    \qquad \text{(Eq.~\ref{eq6})},
    \]
    evaluate the pressure-gradient term using central differencing, and compute the diffusion contribution using the effective viscosity
    \[
    \nu_{\mathrm{eff},i}
=
\max\left(\nu-\nu_{\mathrm{num},i},\,\nu_{\min}\right),
\qquad i \in \{x,y\}, \text{(Eq.~\ref{eq21})}.
    \]

    \item Evaluate the diagnostic ratios
    \[
    r_{\nu} \quad \text{(Eq.~\ref{eq9})}, \qquad
    r_{\mathrm{divmom}} \quad \text{(Eq.~\ref{eq10})}, \qquad
    r_{\mathrm{adv}} \quad \text{(Eq.~\ref{eq11})}, \qquad
    r_{\mathrm{num}} \quad \text{(Eq.~\ref{eq12})},
    \]
    and compute the normalized controller variables
    \[
    z_{\mathrm{num}} \quad \text{(Eq.~\ref{eq15})}, \qquad
    z_{\mathrm{adv}} \quad \text{(Eq.~\ref{eq16})}, \qquad
    z_{\mathrm{div}} \quad \text{(Eq.~\ref{eq17})}.
    \]

    \item Construct the baseline blending target
    \[\beta_{\mathrm{target}} \;=\;  a_0
+ a _1\,\sigma(z_{\mathrm{num}})
+ a_2\,\sigma(z_{\mathrm{adv}}) \\
 - a_3\,\sigma(z_{\mathrm{div}})
+ a_4\,t,\text{(Eq.~\ref{eq18})},
    \]
    then apply regime-based corrections for divergence-limited, over-diffused, and convection-ready states
    \[
    \beta_{\mathrm{target}}
    \qquad \text{(Eq.~\ref{eq19})},
    \]
    and update the blending factor through the relaxed recursion
    \[
    \beta^{updated}
=
(1-\alpha_{\beta})\beta^{previous}
+\alpha_{\beta}\beta_{\mathrm{target}} \text{(Eq.~\ref{eq20})}.
    \]

    \item If projection is activated, correct the provisional velocity field through
    \[
    \mathbf{u}^{\mathrm{corr}}=\mathbf{u}^{*}-\nabla \phi
    \qquad \text{(Eq.~\ref{eq22})},
    \]
    where $\phi$ is obtained from
    \[
    \nabla^2\phi=\nabla\cdot\mathbf{u}^{*}
    \qquad \text{(Eq.~\ref{eq23})}.
    \]

    \item Assemble the component-wise momentum residuals
    \[
    \mathbf{R}_{\mathrm{mom}}
    =
    (\mathbf{u}\cdot\nabla)\mathbf{u}
    +\nabla p
    -\frac{1}{Re}\nabla^2\mathbf{u}
    \qquad \text{(Eq.~\ref{eq24})},
    \]
    and the divergence residual
    \[
    \mathbf{R}_{\mathrm{div}}=\nabla\cdot\mathbf{u}
    \qquad \text{(Eq.~\ref{eq25})}.
    \]

    \item Construct the loss terms
    \[
    L_{\mathrm{mom}}
    \qquad \text{(Eq.~\ref{eq26})}, \qquad
    L_{\mathrm{div}}
    \qquad \text{(Eq.~\ref{eq27})}, \qquad
    L_{\mathrm{total}}
    =
    \lambda_{\mathrm{div}}L_{\mathrm{div}}+\lambda_{\mathrm{mom}}L_{\mathrm{mom}}
    \qquad \text{(Eq.~\ref{eq28})},
    \]
    with epoch-dependent weights
    \[
    \lambda_{\mathrm{mom}}=0.20+0.80t^2,
    \qquad
    \lambda_{\mathrm{div}}=20(1-t)^2+2
    \qquad \text{(Eq.~\ref{eq29})}.
    \]

    \item Minimize $L_{\mathrm{total}}$ by backpropagation using Adam,
    \[
    \theta \leftarrow \theta-\eta_s\nabla_{\theta}L_{\mathrm{total}},
    \]
    while monitoring the residuals and controller diagnostics. If the stage plateaus, reduce the learning rate and extend the stage if necessary.

    \item Perform L-BFGS refinement, followed by sequential variable-refinement steps in which pressure is first held fixed while the velocity components are optimized, then velocity is held fixed while pressure is refined, and finally a coupled optimization is performed for all variables.

    \item Save the converged model for the current stage, transfer the weights to the next finer grid or next Reynolds number, and repeat until all stages and Reynolds numbers are completed.

\end{enumerate}

\vspace{0.1cm}
\hrule

\end{minipage}
\end{center}

\endgroup
\clearpage

\clearpage
\onecolumn

\section*{Notation}
\addcontentsline{toc}{section}{Notation}

\small
\renewcommand{\arraystretch}{1.08}
\setlength{\tabcolsep}{6pt}

\begin{longtable}{
    >{\centering\arraybackslash}p{0.18\textwidth}
    >{\raggedright\arraybackslash}p{0.76\textwidth}
}

\toprule
\textbf{Notation} & \textbf{Description} \\
\midrule
\endfirsthead

\toprule
\textbf{Notation} & \textbf{Description} \\
\midrule
\endhead

\midrule
\multicolumn{2}{r}{Continued on the next page} \\
\endfoot

\bottomrule
\endlastfoot

$\mathcal{A}$
&
Blended discretization of the nonlinear advection component
in the Navier--Stokes momentum equations.
\\

$\mathcal{A}_{\mathrm{central}}$
&
Advection component evaluated using central differencing.
\\

$\mathcal{A}_{\mathrm{upwind}}$
&
Advection component evaluated using the first-order upwind
discretization.
\\

$a_0,\ldots,a_4$
&
Logistic-regression coefficients used to construct the adaptive
blending-factor target, with
$(a_0,a_1,a_2,a_3,a_4)
=(0.05,0.55,0.35,0.45,0.15)$.
\\

$\mathbf b^{(l)}$
&
Trainable bias vector associated with neural-network layer $l$.
\\

$\mathrm{BCs}$
&
Boundary conditions imposed on the predicted flow variables.
\\

$c_x,c_y$
&
Bounded correction factors controlling the retained numerical
viscosity in the $x$- and $y$-coordinate directions, respectively,
where $c_x,c_y\in[0,1]$.
\\

$d$
&
Number of spatial dimensions, where $d=2$ or $3$.
\\

$E$
&
Total number of training epochs assigned to the current multigrid
stage.
\\

$E_{\phi,i}^{\mathrm{abs}}$
&
Pointwise absolute difference between the DNS and NeuralFlowNet
predictions of variable $\phi$ at evaluation point $i$.
\\

$E_s$
&
Number of training epochs prescribed for multigrid stage $s$.
\\

$ep$
&
Current training epoch within a multigrid stage.
\\

$f_{\boldsymbol{\theta}}$
&
Neural-network mapping from the input space $\mathcal X$ to the
output space $\mathcal Y$.
\\

$g^{(l)}$
&
Activation function used in neural-network layer $l$.
\\

$i,j$
&
Spatial-point, data-point, coordinate-direction, or derivative
indices.
\\

$l$
&
Neural-network layer index.
\\

$L_b,L_i,L_r,L_d$
&
Boundary-condition, initial-condition, governing-equation, and
optional observation-data losses, respectively, in the general
PINN formulation.
\\

$L_{\mathrm{div}}$
&
Mean-squared divergence or continuity-residual loss.
\\

$L_{\mathrm{mom}}$
&
Momentum-equation loss constructed from the component-wise
momentum residuals.
\\

$L_{\mathrm{PINN}}$
&
Total loss in the general classical PINN formulation.
\\

$L_{\mathrm{total}}$
&
Total NeuralFlowNet loss combining the divergence and momentum
losses.
\\

$m$
&
Reynolds-number case index in the continuation sequence
$\{Re_m\}$.
\\

$N$
&
Total number of spatial evaluation points used to calculate the
NRMSE.
\\

$N_\Omega$
&
Number of computational points in domain $\Omega$.
\\

$N_s$
&
Computational-grid specification or grid resolution for multigrid
stage $s$.
\\

$\mathcal N_{\boldsymbol{\theta}}$
&
Multilayer-perceptron predictor parameterized by
$\boldsymbol{\theta}$.
\\

$\mathrm{NRMSE}$
&
Normalized root-mean-square error used to evaluate NeuralFlowNet
predictions of $\phi\in\{u,v,w,p\}$ against DNS reference
solutions.
\\

$p,\hat p$
&
Pressure field and pressure predicted by the neural network,
respectively.
\\

$\mathbf q,\hat{\mathbf q}$
&
Primitive-variable vector and its neural-network prediction,
respectively, with
$\hat{\mathbf q}=[\hat u,\hat v,\hat w,\hat p]$.
\\

$R$
&
Solution regime identified by the adaptive controller:
divergence-limited, over-diffused, or advection-ready.
\\

$R_{\mathrm{div}}$
&
Divergence or continuity residual,
$R_{\mathrm{div}}=\nabla\cdot\mathbf u$.
\\

$R_{\mathrm{mom}}$
&
Momentum-residual vector assembled from the advection,
pressure-gradient, and viscous-diffusion contributions.
\\

$R_u,R_v$
&
Residuals of the $x$- and $y$-momentum equations, respectively.
\\

$Re$
&
Reynolds number; the nondimensional physical viscosity is
represented as $\nu=1/Re$.
\\

$Re_m$
&
Reynolds number associated with continuation case $m$.
\\

$r_{\mathrm{adv}}$
&
Advection-strength ratio measuring advection relative to physical
diffusion.
\\

$r_{\mathrm{adv,ready}}$
&
Reference value marking the onset of advection-dominant behavior;
$r_{\mathrm{adv,ready}}=1.20$.
\\

$r_{\mathrm{divmom}}$
&
Divergence-to-momentum ratio comparing the root-mean-square
divergence and momentum residuals.
\\

$r_{\mathrm{div,mid}}$
&
Reference divergence-to-momentum ratio used by the adaptive
controller; $r_{\mathrm{div,mid}}=0.45$.
\\

$r_{\mathrm{instantaneous}}$
&
Current unsmoothed value of a diagnostic ratio.
\\

$r_{\mathrm{num}}$
&
Numerical-diffusion ratio comparing estimated artificial
numerical viscosity with physical viscosity.
\\

$r_{\mathrm{num,target}}$
&
Nominal reference value of the numerical-diffusion ratio;
$r_{\mathrm{num,target}}=0.10$.
\\

$r_{\mathrm{previous}},r_{\mathrm{updated}}$
&
Diagnostic-ratio values before and after exponential smoothing,
respectively.
\\

$r_\nu$
&
Diffusion-dominance ratio measuring physical diffusion relative
to advection and the pressure gradient.
\\

$S$
&
Backward-facing-step height used to nondimensionalize the BFS
geometry and reattachment distance.
\\

$s$
&
Multigrid training-stage index.
\\

$t$
&
Normalized epoch coordinate,
$t=ep/\max(E-1,1)$.
\\

$\mathbf u$
&
Velocity vector, with $\mathbf u=(u,v)$ in 2D and
$\mathbf u=(u,v,w)$ in 3D.
\\

$\mathbf u^*,\mathbf u_{\mathrm{corr}}$
&
Provisional velocity and projection-corrected velocity fields,
respectively.
\\

$u,v,w$
&
Streamwise, transverse or wall-normal, and spanwise velocity
components, respectively.
\\

$\hat u,\hat v,\hat w$
&
Velocity components predicted by the neural network.
\\

$U_b$
&
Bulk velocity used for normalization in the periodic-hill and
backward-facing-step cases; $U_b=1$.
\\

$|\mathbf U|,U_{\mathrm{mag}}$
&
Velocity magnitude.
\\

$\mathbf W^{(l)}$
&
Trainable weight matrix associated with neural-network layer $l$.
\\

$\mathbf x$
&
Neural-network input vector containing spatial coordinates.
\\

$x,y,z$
&
Streamwise, transverse or wall-normal, and spanwise spatial
coordinates, respectively.
\\

$x_i,x_j$
&
Spatial coordinates in directions $i$ and $j$, respectively.
\\

$x/S$
&
Streamwise distance normalized by the backward-facing-step
height.
\\

$\mathcal X,\mathcal Y$
&
Neural-network input and output spaces, respectively.
\\

$\mathbf z^{(l)}$
&
Output or activation vector of neural-network layer $l$.
\\

$z_{\mathrm{adv}}$
&
Standardized advection-strength diagnostic.
\\

$z_{\mathrm{div}}$
&
Standardized divergence-to-momentum diagnostic.
\\

$z_{\mathrm{num}}$
&
Standardized numerical-diffusion diagnostic.
\\

$\alpha$
&
Relaxation coefficient used to smooth both the diagnostic ratios
and the adaptive blending factor; $\alpha=0.08$.
\\

$\beta$
&
Adaptive factor blending the upwind and central advection
discretizations, where $0\leq\beta\leq1$.
\\

$\beta^{\mathrm{previous}},
 \beta^{\mathrm{updated}}$
&
Blending factors before and after the relaxed controller update,
respectively.
\\

$\beta_{\mathrm{target}}$
&
Target blending factor calculated from the diagnostic ratios,
training progress, and detected solution regime.
\\

$\Delta x,\Delta y$
&
Local computational-grid spacings in the $x$- and
$y$-coordinate directions, respectively.
\\

$\epsilon$
&
Small positive constant added to denominators to prevent division
by zero.
\\

$\eta_s$
&
Learning rate assigned to multigrid stage $s$.
\\

$\boldsymbol{\theta},
 \boldsymbol{\theta}^*$
&
Trainable and converged neural-network parameters, respectively.
\\

$\lambda_b,\lambda_i,\lambda_r,\lambda_d$
&
Weights assigned to the boundary-condition, initial-condition,
PDE-residual, and optional observation-data losses, respectively.
\\

$\lambda_{\mathrm{div}},
 \lambda_{\mathrm{mom}}$
&
Epoch-dependent weights assigned to the divergence and momentum
losses, respectively.
\\

$\lambda_u,\lambda_v$
&
Weights assigned to the $x$- and $y$-momentum residuals,
respectively; both are set to $1.0$.
\\

$\nu$
&
Physical kinematic viscosity; represented as $\nu=1/Re$ in the
nondimensional formulation.
\\

$\nu_{\mathrm{eff},i}$
&
Effective directional viscosity used in the discrete diffusion
operator.
\\

$\nu_{\min}$
&
Minimum effective viscosity retained for numerical stability,
approximately $1\%$--$3\%$ of the physical viscosity.
\\

$\nu_{\mathrm{num},i}$
&
Estimated artificial numerical viscosity in coordinate direction
$i$.
\\

$\nu_{\mathrm{num},x},
 \nu_{\mathrm{num},y}$
&
Estimated artificial numerical viscosities in the $x$- and
$y$-coordinate directions, respectively.
\\

$\rho$
&
Fluid density.
\\

$\sigma(\cdot)$
&
Sigmoid function used to map standardized diagnostics to bounded
controller responses.
\\

$\phi$
&
Generic flow variable used in the validation metrics, where
$\phi\in\{u,v,w,p\}$.
\\

$\phi_{\mathrm{DNS},i},
 \phi_{\mathrm{NFN},i}$
&
DNS and NeuralFlowNet values, respectively, of variable $\phi$
at evaluation point $i$.
\\

$\phi_{\mathrm{DNS,max}},
 \phi_{\mathrm{DNS,min}}$
&
Maximum and minimum DNS values of variable $\phi$ over the
evaluation domain.
\\

$\varphi$
&
Scalar correction potential used in the projection-based velocity
correction.
\\

$\Omega$
&
Computational domain.
\\

$\|\cdot\|$
&
Root-mean-square norm used in the adaptive diagnostic ratios.
\\

$\nabla,\nabla\cdot,\nabla^2$
&
Gradient, divergence, and Laplacian operators, respectively.
\\

$\partial/\partial t$
&
Temporal-derivative operator included in the general
Navier--Stokes equations but omitted from the steady-state
NeuralFlowNet residuals.
\\

$\operatorname{mean}(\cdot)$
&
Arithmetic mean evaluated over the computational domain.
\\

\end{longtable}

\normalsize

\setcounter{figure}{0}
\renewcommand{\thefigure}{A\arabic{figure}}

\begin{figure*}[!h]
    \centering

    \includegraphics[width=1\textwidth]
    {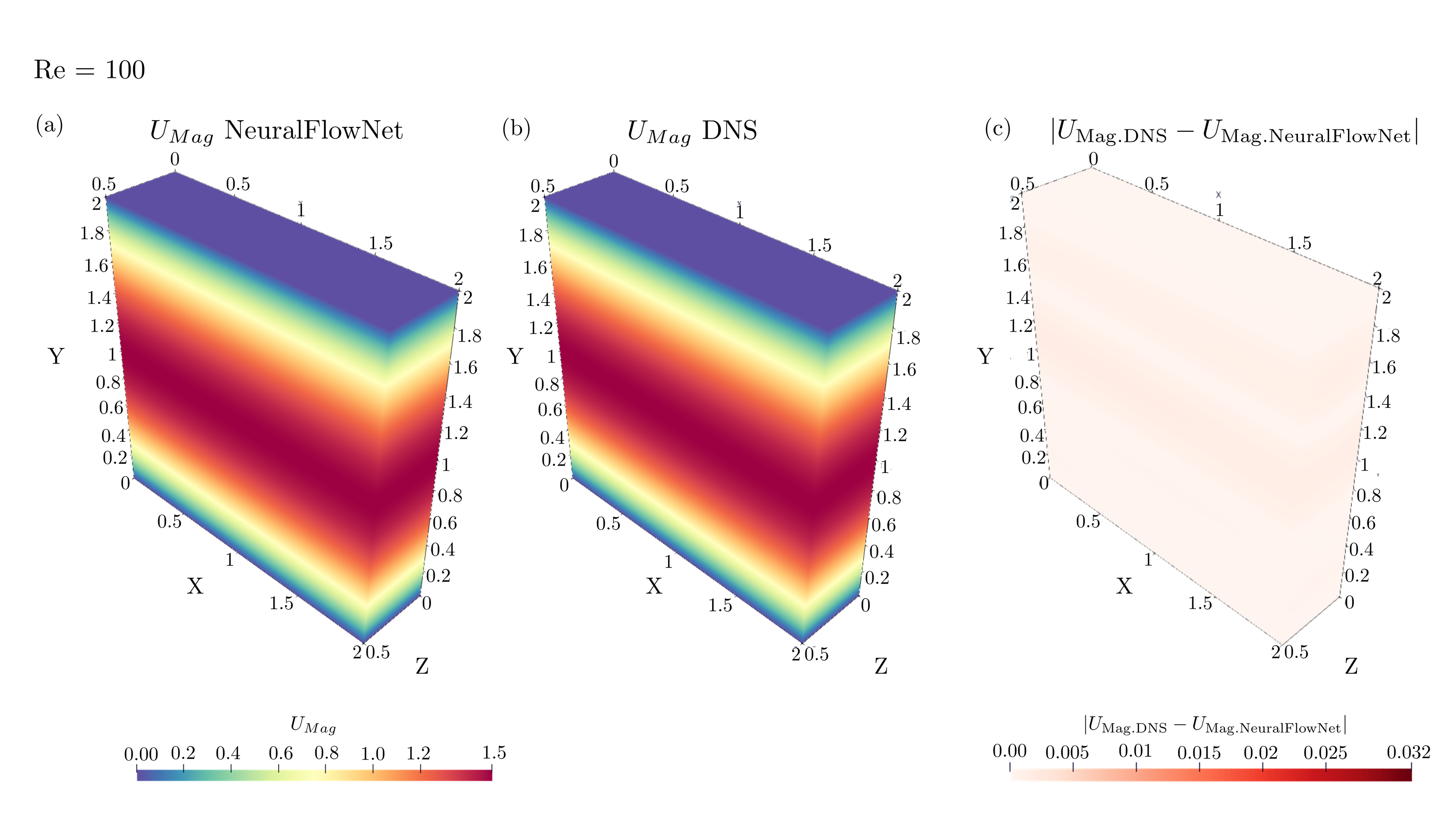}

    \vspace{0.1cm}

    \includegraphics[width=1\textwidth]
    {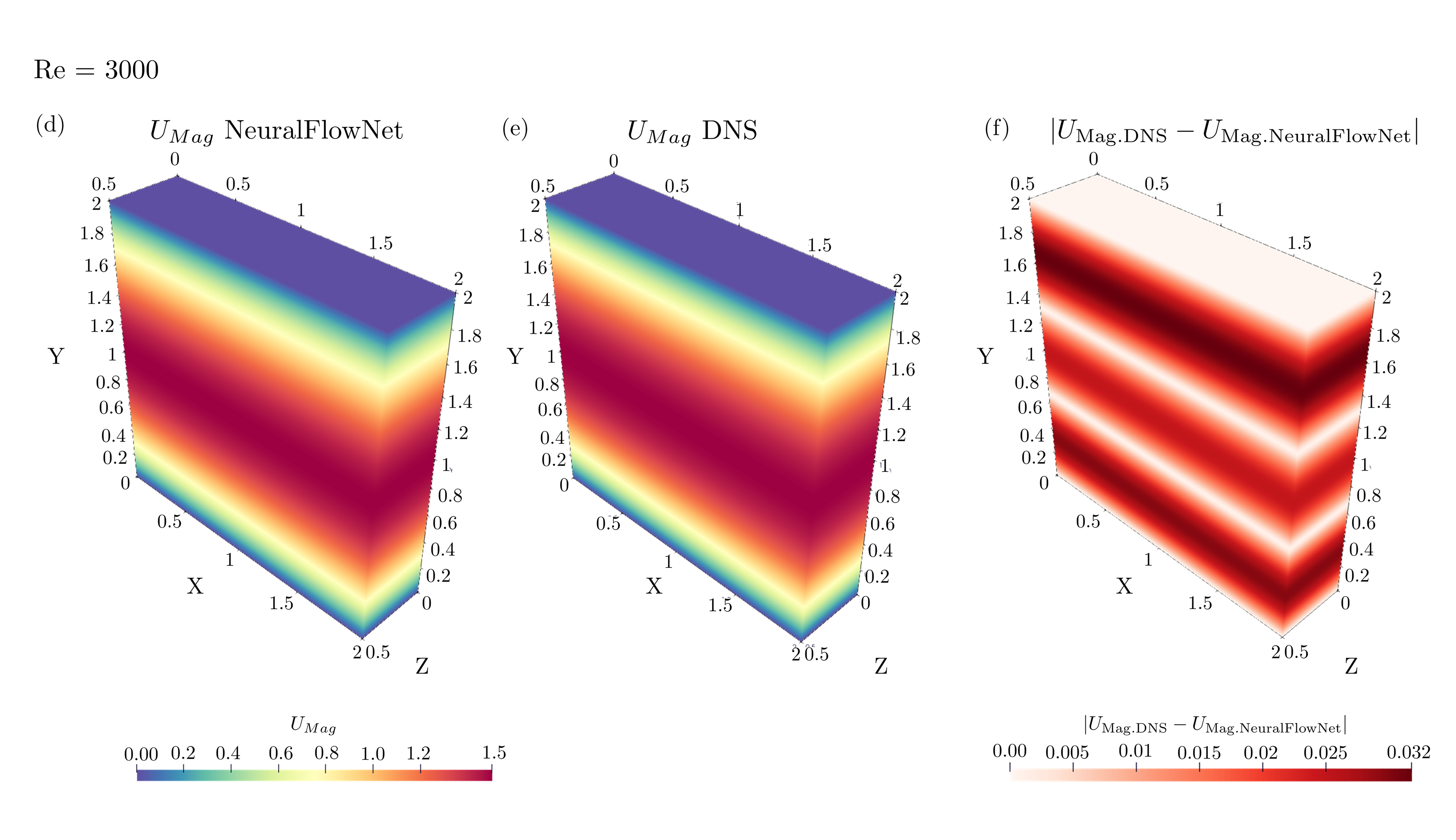}

    \caption{Comparison of 3D internal-flow predictions by
    \textit{NeuralFlowNet} and the reference solution at
    $Re=100$ and $Re=3000$. Panels (a)--(c) show
    $U_{\mathrm{mag}}$ from \textit{NeuralFlowNet}, the
    reference solution, and the absolute difference for $Re=100$,
    respectively. The same comparison is repeated for
    $Re=3000$ in the remaining panels.}

    \label{fig:appendix_pipe}
\end{figure*}

\end{document}